\documentclass[a4paper,twocolumn,11pt]{quantumarticle}
\pdfoutput=1
\usepackage[utf8]{inputenc}
\usepackage[english]{babel}
\usepackage[T1]{fontenc}
\usepackage{amsmath}
\usepackage{hyperref}
\usepackage{tikz}
\usepackage{lipsum}
\usepackage[numbers,sort&compress]{natbib}

\usepackage{graphicx}
\usepackage{dcolumn}
\usepackage{bm}
\usepackage{amssymb,blkarray}
\usepackage{color,soul}
\usepackage{multirow}

\begin{document}

\title{Genuine tripartite entanglement in a  mixed spin-(1/2,1) Heisenberg tetramer}

\author{Hana  Vargov\'a}
\email{hcencar@saske.sk}
\affiliation{Institute of Experimental Physics, Slovak Academy of Sciences, Watsonova 47, 040 01 Ko\v {s}ice, Slovakia}
\orcid{0000-0003-2000-7536}
\author{Jozef Stre\v{c}ka}
\orcid{0000-0003-1667-6841}
\affiliation{Department of Theoretical Physics and Astrophysics, Faculty of Science, Pavol  Jozef \v{S}af\'{a}rik University, Park 
Angelinum 9, 040 01 Ko\v{s}ice, Slovakia}

\maketitle

\begin{abstract}
A genuine tripartite entanglement of a mixed spin-(1/2,1) Heisenberg tetramer is rigorously analyzed in a presence of external magnetic field. The couple of mixed spin-(1/2,1) dimers is arranged in a perfect rectangular square plaquette involving two nonequivalent Heisenberg exchange couplings $J$ and $J_1$. The degree of a genuine tripartite entanglement is evaluated  according to the genuine tripartite negativity ${\cal N}_{ABC}$ defined as a geometric mean of all possible bipartite negativities corresponding to a decomposition into a single spin and the remaining spin dimer ${\cal N}_{A|BC}$, ${\cal N}_{B|AC}$ and ${\cal N}_{C|AB}$ after degrees of freedom of the last fourth spin $D$ are traced out. Due to the symmetry of a mixed spin-(1/2,1) Heisenberg tetramer two different genuine tripartite negativities for the trimeric system $1/2\!-\!1\!-\!1$  and $1/2\!-\!1/2\!-\!1$ were identified. It was found that the genuine tripartite negativity for the interaction ratio $J_1/J\!<\!1$ becomes nonzero solely in the tripartite system $1/2\!-\!1\!-\!1$ at low-enough magnetic fields. The opposite interaction limit $J_1/J\!>\!1$  gives rise to the nonzero genuine tripartite negativity in both tripartite systems in a presence of external magnetic field until the classical ferromagnetic state is achieved. It was shown, that the genuine tripartite negativity of a mixed spin-(1/2,1) Heisenberg tetramer can be detected also at nonzero temperatures. An enhancement of the thermal genuine tripartite negativity through the enlargement of the total spin number of a tripartite system is evidenced. The correlation between the bipartite negativity of two spins and the genuine tripartite negativity is discussed in detail. 
\end{abstract}

\section{Introduction}
\label{introduction}

The molecular magnetism is nowadays one of the rapidly progressing research area connecting the state-of-the-art knowledge from the quantum physics, quantum chemistry, condensed matter physics and/or computational science. Enormous activities in the field of molecular magnetism originate from the nontrivial features of these low-dimensional complexes, whose utilization in various devices get started a new era of technical applications. Among these one could for example mention the manipulation of magnetism of each individual molecule in a complex paramagnetic salt as a potential platform for an ultra-high-density storage device, in which one molecule could store a one bit of information~\cite{Liu15}. Another specific phenomenon is  presence of a quantum tunneling of the magnetization below the blocking temperature, which may become a key ingredient for a quantum computing~\cite{Abel}. The molecular cooling is an additional peculiar physical phenomenon observed in   molecular magnets~\cite{Sharples}, which allows us to achieve  very low temperatures. Last but not least, the inherence of  nonlocal correlations in molecular magnets makes them very attractive for a quantum cryptography, quantum teleportation of information as well as the quantum computing~\cite{Deutsch,Furusawa,Loss}.

Due to the  nonlocal correlations between two components of an entangled pair, it is completely sufficient to know the state of only the one of them and the state of the remaining one is unambiguously determined without its measurement.  In general, the composite quantum system is said to be entangled, if a quantum state of  selected  component of a composite system cannot be described independently of the state of the other component. As was shown in various experimental and theoretical studies the entanglement can be induced, manipulated and destroyed, see Ref.~\cite{Amico,HorodeckiRev} for reviews and references therein. In the literature one can find a great variety of  classifications of an entanglement,  e.g., the zero-temperature  and the thermal entanglement, the bipartite and the multipartite entanglement, the genuine entanglement and the entanglement between the single spin and the remaining spin multiplet. Of course, each of these classifications requires an appropriate quantification procedure to better understand the underlying physics.  A significant progress in a comprehension of an entanglement allows us to quite well understand all aspects of bipartite entanglement between two  entities, however, the understanding of multipartite entanglement is still incomplete. Due to the calculation complexity of a multipartite problem, up to now, there exist only a few theoretical studies of smaller quantum Heisenberg systems such as trimers~\cite{Benabdallah22,Tribedi,Wang2001,Bose,Pal2009,Pal2011,Cima,Liu2015, Deniz, Ahami, Zhang2021,Najarbashi,Zad2017, Zad2016,Zheng,Sun}, tetramers~\cite{Karlova23,Vargova2023,Sun,Tribedi,Irons,Bose,Pal2011,Benabdallah,Szalowski,Torrico,Rojas, Ananikian,Shawish,Ma,Zhad2022,Li,Zheng2017,Abgaryan2015,Hu,Karlova,Ghannadan2021}, pentamers~\cite{Sun,Xu2016,Szalowski2023}, hexamers~\cite{Deb,Zhad2018} or even larger structures~\cite{Wang2002,Sadiek,Sadiek2}.   In the present article we would like to bring a deeper  insight into the problem of multipartite entanglement and we would like to continuously  link to our previous study dealing with the distribution of a bipartite entanglement between two  spins in a mixed spin-(1/2,1) Heisenberg tetramer with a geometry of a square plaquette~\cite{Vargova2023}.  In particular, we would like to analyze the genuine tripartite entanglement within a tetrapartite system through  the behavior of negativity, which was conceptually introduced by Peres~\cite{Peres} and Horodecki~\cite{Horodecki}. They independently postulated that all eigenvalues of a partially transposed density matrix of each separable bipartite state should be  non-negative. It should be emphasized that the original definition of a bipartite negativity allows one to quantify the bipartite entanglement between two entities. In case of a multipartite system, the original definition could be slightly modified and all bipartite, tripartite and multipartite negativities in precisely defined hierarchy should be analyzed~\cite{Dur2000}.  In the spirit of this fact, our analysis of a genuine tripartite negativity relies on the behavior of all possible bipartite negativities between a single spin, say $A$, and the  remaining spin pair, say $B\!-\!C$, if degrees of freedom of the last four spin, say $D$, from the square plaquette are traced out~\cite{Vidal}.   Beside this, we will try to investigate  correlation between a genuine tripartite negativity of a mixed spin-(1/2,1) Heisenberg tetramer and the respective bipartite negativities of two selected  spins~\cite{Vargova2023}.

The structure of the paper is organized as follows. In section~\ref{model} we will introduce the  quantum spin model under the investigation and describe basic steps of the calculation procedure. In section~\ref{results} we will report the most interesting results for the measures of bipartite entanglement between an individual spin and remaining spin dimer as well as the behavior of a genuine tripartite entanglement. The stability of entanglement at nonzero temperatures will be also discussed in detail.  Finally, the section~\ref{conclusion} provides a brief summary of the most important scientific findings. Some technical details of calculations  are  summarized in Appendices~\ref{App A}-\ref{App F}.
\section{Model and Method}
\label{model}
\begin{figure}[h]
\centering
{\includegraphics[width=.18\textwidth,trim=1.0cm 4.5cm 16cm 20.5cm, clip]{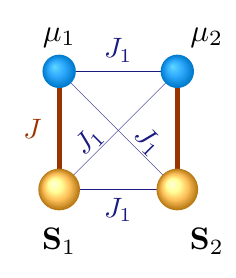}}
\caption{A schematic representation of the mixed spin-(1/2,1) Heisenberg tetramer with a geometric structure of a square plaquette. Small (blue) and large (yellow) balls correspond to spins  $S_{i}\!=\!1$ and $\mu_{i}\!=\!1/2$, respectively. Four spins interact between themselves via the Heisenberg coupling $J$ (thick vertical lines) and $J_1$ (thin horizontal and diagonal lines).}
\label{fig1}
\end{figure}
The Hamiltonian of a mixed spin-(1/2,1) Heisenberg tetramer can be expressed as follows:
\allowdisplaybreaks
\begin{align}
\hat{\cal H}&=J\left(\hat{\bf S}_{1}\!\cdot\!\hat{\boldsymbol\mu}_{1}
\!+\!\hat{\bf S}_{2}\!\cdot\!\hat{\boldsymbol\mu}_{2}\right)\!+\! J_1\left(\hat{\bf S}_{1}\!+\!\hat{\boldsymbol\mu}_{1}\right)\!\cdot\!\left(\hat{\bf S}_{2}\!+\!\hat{\boldsymbol\mu}_{2}\right)
\nonumber\\
\!&-\!h\left(\hat{S}^z_{1}\!+\!\hat{S}^z_{2}\!+\!\hat{\mu}^z_{1}\!+\!\hat{\mu}^z_{2}\right).
\label{eq1}
\end{align} 
Two types of  spins  ${\bf S}_i\!=\!1$ and ${{{\boldsymbol\mu}}_i}\!=\!1/2$ interact between themselves via the  isotropic Heisenberg interactions $J$ and $J_1$  in concordance with the scheme presented in Fig.~\ref{fig1}. The influence of external magnetic field $B$ is described by the parameter $h\!=\!g\mu_BB$, where 
$g$ is a  Land\'e $g$-factor identical for both spins and $\mu_B$ is a Bohr magneton.
 
In order to analyze the genuine tripartite entanglement among the $A$, $B$ and $C$ entities of a mixed spin-(1/2,1) Heisenberg tetrapartite system $A-B-C-D$ we utilize  the definition  of a genuine tripartite negativity~\cite{Sabin}
\begin{align}
{\cal N}_{ABC}\!=\!\left( {\cal N}_{A|BC} {\cal N}_{B|AC} {\cal N}_{C|AB} \right)^{1/3},
\label{eq2}
\end{align}
which is given by a geometric mean of  bipartite negativities between a single   spin, say $A$, and the remaining two spins, say $B\!-\!C$, after tracing out degrees of freedom of the forth spin, say $D$. Such bipartite negativity measuring the entanglement between a single spin $A$ and the spin dimer $B\!-\!C$ is denoted as ${\cal N}_{A|BC}$. Of course, the knowledge of all further permutations is required.
To evaluate the bipartite negativity ${\cal N}_{A|BC}$ we will adapt the definition introduced by Vidal and Werner~\cite{Vidal} from which the bipartite negativity ${\cal N}_{A|BC}$  should be calculated as  a sum of absolute values of all negative  eigenvalues $(\lambda_{A|BC})_i$ of a partially transposed density matrix $\hat{\rho}_{A|BC}^{T_{A}}$. The transposition is performed over  the  subsystem $A$, which is denoted by the upper superscript $T_A$. Hence, 
\begin{align}
{\cal N}_{A|BC}\!=\!{\cal N}(\rho_{{A|BC}})\!=\!\sum_{(\lambda_{A|BC})_i<0}\left\vert(\lambda_{A|BC})_i\right\vert.
\label{eq3}
\end{align}
The respective density  operator $\hat{\rho}_{A|BC}$ in case of a mixed spin-(1/2,1) Heisenberg tetramer is calculated  by tracing out the degrees of freedom of the remaining spin $D$,
\begin{align}
\hat{\rho}_{A|BC}\!&=\!
\frac{1}{\cal Z}\sum_{k=1}^{36} {\rm Tr}_{D}{\rm e}^{-\beta \varepsilon_k}\vert \psi_k\rangle \langle \psi_k\vert,
\label{eq4}
\end{align}
where the summation runs over all possible 36 eigenvectors $\vert \psi_k\rangle$ with the respective eigenvalues $\varepsilon_k$ of the mixed spin-(1/2,1) Heisenberg tetramer given by the Hamiltonian~\eqref{eq1} and the symbol  ${\cal Z}$ denotes the partition function. It is worthwhile to remark that a detailed analysis of the ground-state properties as well as the  explicit expressions of all relevant quantities can be found in our previous paper~\cite{Vargova2023}.

Due to the underlying symmetry the mixed spin-(1/2,1) Heisenberg tetramer~\eqref{eq1} with the magnetic structure of a square plaquette  schematically shown in Fig.~\ref{fig1} is characterized by two different genuine  tripartite negativities
\begin{align}
{\cal N}_{\mu_{1}S_{1}S_{2}}\!&=\!\left( {\cal N}_{\mu_{1}|S_{1}S_{2}} {\cal N}_{S_{1}|\mu_{1}S_{2}} {\cal N}_{S_{2}|\mu_{1}S_{1}} \right)^{1/3},
\nonumber\\
{\cal N}_{\mu_{1}\mu_{2}S_{1}}\!&=\!\left( {\cal N}_{\mu_{1}|\mu_{2}S_{1}} {\cal N}_{\mu_{2}|\mu_{1}S_{1}} {\cal N}_{S_1|\mu_1\mu_{2}} \right)^{1/3},
\label{eq5}
\end{align}
which are defined through the six different bipartite negativities ${\cal N}_{\mu_{1}|S_{1}S_{2}}$, ${\cal N}_{S_{1}|\mu_{1}S_{2}}$, $ {\cal N}_{S_{2}|\mu_{1}S_{1}}$, ${\cal N}_{\mu_{1}|\mu_{2}S_{1}}$, ${\cal N}_{\mu_{2}|\mu_{1}S_{1}}$ and  ${\cal N}_{S_1|\mu_1\mu_{2}}$.  A detailed derivation of each aforementioned bipartite negativity is given in Appendices~\ref{App A} to \ref{App F}.  It is quite obvious from its definition that the magnitude of a genuine tripartite negativity always achieves a non-negative value. If the magnitude of a genuine tripartite negativity is zero, then  at least one bipartite negativity ${\cal N}_{A|BC}$ or ${\cal N}_{B|AC}$ or ${\cal N}_{C|AB}$ is zero  and thus, the states of at least one spin should be  separable with respect to the other two. On the other hand, if the magnitude of a genuine tripartite negativity is nonzero  the system  of three spins $A\!-\!B\!-\!C$ is fully inseparable and each bipartite negativity  ${\cal N}_{A|BC}$, ${\cal N}_{B|AC}$ and ${\cal N}_{C|AB}$ must be nonzero.
For the completeness we note  that if a  bipartite negativity  is zero, say ${\cal N}_{A|BC}$, then, the spin $A$ is  separable from the spin dimer $B\!-\!C$, otherwise  the  spin $A$ is  entangled with the spin dimer $B\!-\!C$. 

\section{Results and discussion}
\label{results}
In accordance with our previous analysis~\cite{Vargova2023}, we restricted our attention only to the physically most interesting case with both  interactions being antiferromagnetic ($J\!>\!0$, $J_1\!\geq\!0$). In the following  all quantities will be  normalized with respect to the coupling constant $J$.
\begin{figure*}[t!]
\centering
{\includegraphics[width=.45\textwidth,trim=1.8cm 8.45cm 0.5cm 8cm, clip]{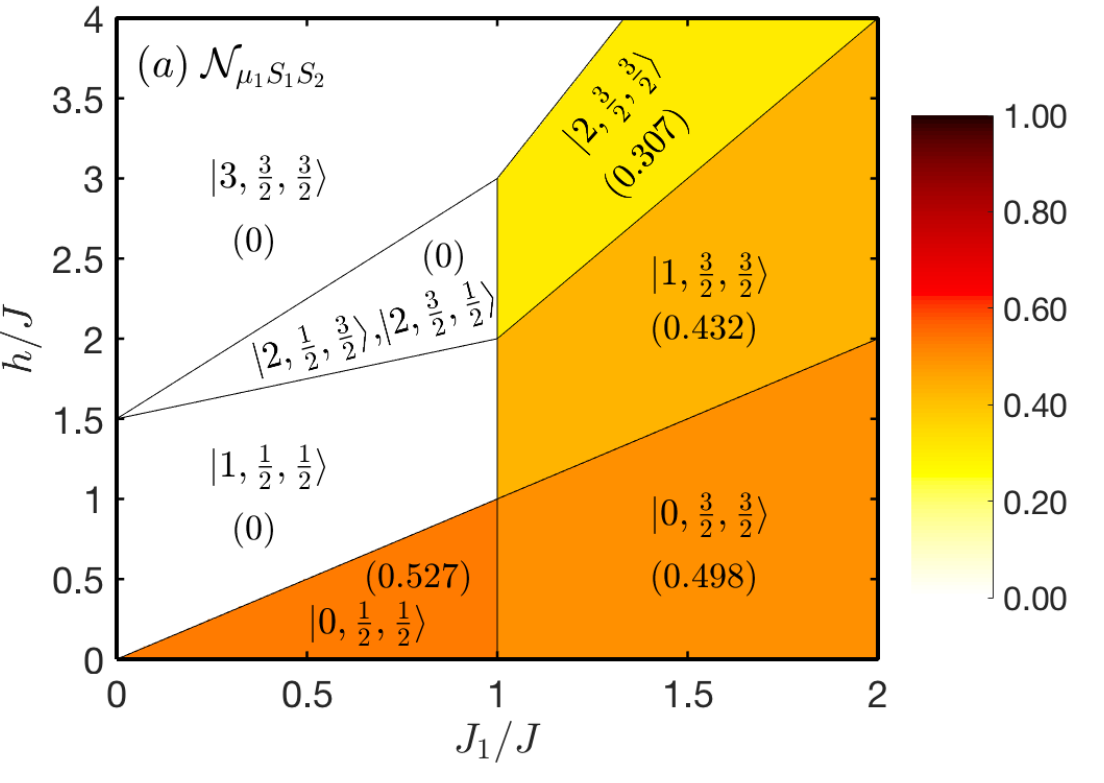}}
{\includegraphics[width=.45\textwidth,trim=1.8cm 8.45cm .5cm 8cm, clip]{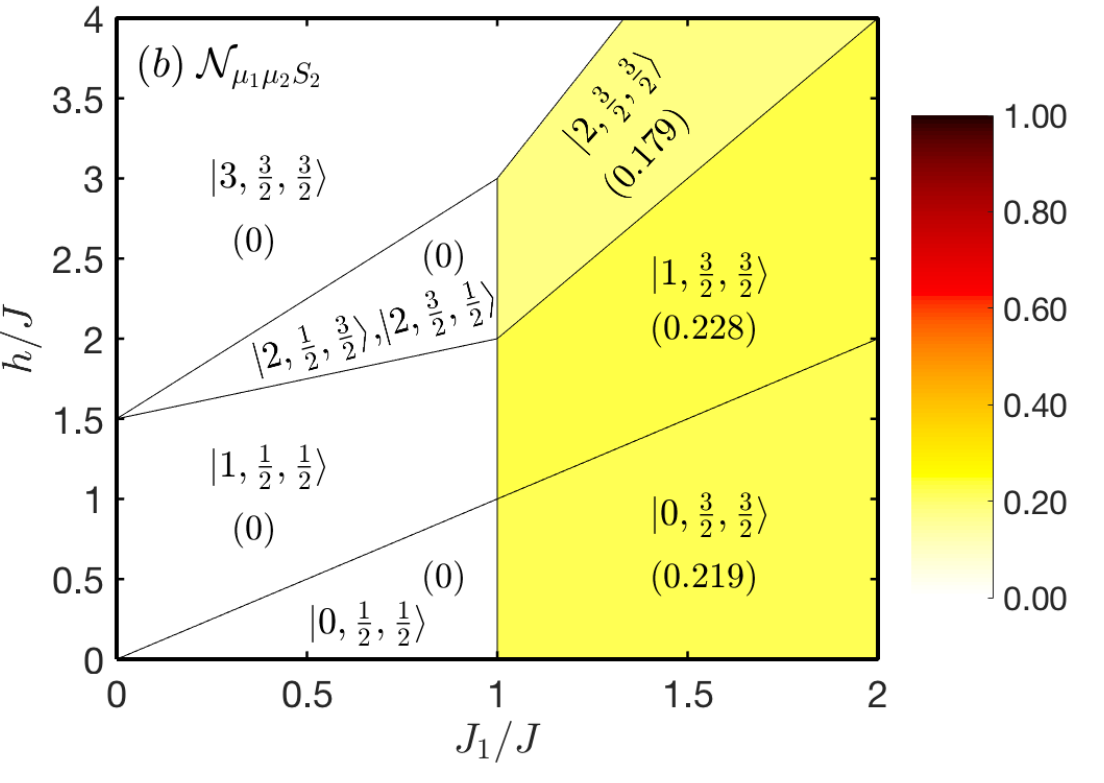}}
\caption{Density plots of  quantum genuine tripartite negativities ${\cal N}_{\mu_1S_1S_2}$ ($a$) and ${\cal N}_{\mu_1\mu_2S_2}$ ($b$) of the mixed spin-(1/2,1) Heisenberg tetramer in the $J_1/J-h/J$ plane. Solid black lines determine phase boundaries between different magnetic ground states characterized by a set of three quantum numbers $|\sigma_T^z,\sigma_{1},\sigma_{2}\rangle$. The values in the round brackets denote the magnitude of the genuine tripartite negativity in the respective ground state.}
\label{fig2a}
\end{figure*}

First, we pay attention to an exhaustive analysis of a tripartite entanglement of a mixed spin-(1/2,1) Heisenberg tetramer at zero temperature. As was shown previously~\cite{Vargova2023}, the ground-state phase diagram involves seven different phases characterized by a  set of three quantum numbers $\vert \sigma_T^z, \sigma_{1}, \sigma_{2}\rangle$,   ${\sigma}_T^z$ is a $z$-projection of a total spin operator $\hat{\boldsymbol\sigma}_T\!=\!\sum_{\gamma=1,2}\hat{\boldsymbol\sigma}_{\gamma}\!=\!\sum_{\gamma=1,2}(\mathbf{\hat{S}}_{\gamma}\!+\!\hat{\boldsymbol\mu}_{\gamma})$, with the following values $\sigma_T^z\!=\!-\sigma_T,\dots,\sigma_T$; $\sigma_T\in\{ |\sigma_{1}\!-\!\sigma_{2}|,|\sigma_{1}\!-\!\sigma_{2}|+1,\dots,\sigma_{1}\!+\!\sigma_{2}\}$ and the composite spin operators of two interacting dimers acquire the following eigenvalues $\sigma_{\gamma}\!\in\!\left\{ 1/2, 3/2\right\}$ for $\gamma\!=\!1,2$, respectively.  

The distribution of both types of a   genuine tripartite entanglement quantified through the  negativity within  the ground-state phase diagram in the $J_1/J-h/J$ plane is presented in  Fig.~\ref{fig2a}. 
It is  evident from Fig.~\ref{fig2a} that the degree of  genuine tripartite negativities ${\cal N}_{\mu_1S_1S_2}$ and ${\cal N}_{\mu_1\mu_2S_2}$   is specific for each ground state and is independent of the magnetic field and the interaction ratio $J_1/J$ until  discontinuous changes are reached at the relevant ground-state phase boundaries. The  degrees of  tripartite negativities,  which determine the magnitude of both  genuine tripartite entanglement, ${\cal N}_{\mu_1S_1S_2}$ and  ${\cal N}_{\mu_1\mu_2S_2}$,  are given in Tab.~\ref{tab2}. 
\begin{table*}[t!]
\resizebox{0.98\textwidth}{!}{
\begin{tabular}{l || l || l | l | l | l|l|l }
  $J_1/J$ & $|\sigma^z_T,\sigma_{1},\sigma_{2}\rangle$ & ${\cal N}_{\mu_{1}|S_{1}S_{2}}$ & ${\cal N}_{S_{1}|\mu_{1}S_{2}}$ & ${\cal N}_{S_{2}|\mu_{1}S_{1}}$ & ${\cal N}_{\mu_{1}|\mu_{2}S_{2}}$ & ${\cal N}_{\mu_{2}|\mu_{1}S_{2}}$& ${\cal N}_{S_{2}|\mu_{1}\mu_{2}}$\\\hline\hline
   $<1$ & $\vert 0,\frac{1}{2},\frac{1}{2}\rangle$ & $\frac{1}{18}(\sqrt{97}\!-\!1)\approx 0.492$ &$\frac{1}{18}(4\sqrt{5}\!+\!\sqrt{17}\!+\!3)\approx 0.893$& 1/3$\approx$0.333 & 0 &$\frac{1}{12}(1\!+\!\sqrt{17})\approx$0.427& $\frac{1}{9}(3\sqrt{2}\!-\!1)\approx$0.360\\
         &&  & &  & &&\\
         & $\vert 1,\frac{1}{2},\frac{1}{2}\rangle$ &$\sqrt{2}/3\approx$0.471 & $\sqrt{2}/3\approx$0.471  & 0 & 0&$\sqrt{2}/3\approx$0.471&$\sqrt{2}/3\approx$0.471\\
          &&  & &  & &&\\
       & $\vert 2,\frac{1}{2},\frac{3}{2}\!\rangle$, $\vert 2,\frac{3}{2},\frac{1}{2}\rangle$ &  $\frac{1}{6}(\sqrt{3}\!-\!1)\approx$0.122 &  $\frac{1}{6}(\sqrt{3}\!-\!1)\approx$0.122 & 0 & 0&1/6$\approx$0.167&1/6$\approx$0.167\\
           &&  & &  & &&\\
   & $\vert 3,\frac{3}{2},\frac{3}{2}\rangle$ & 0 & 0 & 0 & 0&0&0\\
           &&  & &  & &&\\
\hline
   $=1$ & $\vert 0,\frac{1}{2},\frac{1}{2}\rangle$, & 1/6$\approx$0.167& 1/2=0.500 & 1/2=0.500 & 1/8=0.125 &1/8=0.125&1/6$\approx$0.167\\
&$\vert 0,\frac{3}{2},\frac{3}{2}\rangle$ &  & &  & &&\\
         &&  & &  & &&\\
         & $\vert 1,\frac{1}{2},\frac{1}{2}\rangle$,  & A$\approx$0.111 & C$\approx$0.205 & C$\approx$0.205 &F$\approx$0.049 & F$\approx$0.049 & H$\approx$0.089\\
         &$\vert 1,\frac{3}{2},\frac{3}{2}\rangle$,  &  & &  & &\\
        &$\vert 1,\frac{1}{2},\frac{3}{2}\rangle$, &    & &  & &&\\
        &$\vert 1,\frac{3}{2},\frac{1}{2}\rangle$  &  & &  & &&\\
         &&  & &  & &&\\
        & $\vert 2,\frac{1}{2},\frac{3}{2}\!\rangle$, $\vert 2,\frac{3}{2},\frac{1}{2}\rangle$, & $\frac{1}{36}(\sqrt{41}\!-\!5)\approx$0.039 & $\frac{1}{18}\approx$0.056 & $\frac{1}{18}\approx$0.056 & $\frac{1}{18}(\sqrt{7}\!-\!2)\approx$0.036&$\frac{1}{18}(\sqrt{7}\!-\!2)\approx$0.036 & $\frac{1}{9}(\sqrt{2}\!-\!1)\approx$0.046\\
& $\vert 2,\frac{3}{2},
 \frac{3}{2}\rangle$ &  & &  & &&\\ 
          &&  & &  & &&\\
    & $\vert 3,\frac{3}{2},\frac{3}{2}\rangle$ & 0 & 0 & 0 & 0&0 & 0\\
            &&  & &  & &&\\
	      \hline	   
	  $>1$ & $\vert 0,\frac{3}{2},\frac{3}{2}\rangle$ & 1/3$\approx$0.333 & $\frac{1}{36}(\sqrt{89}\!+\!4\sqrt{41}\!-\!15)\approx$0.557 & 2/3$\approx$0.666 & 1/4$\approx$0.250&$\frac{1}{24}(\sqrt{89}\!-\!5)\approx$0.185 &$\frac{1}{35}(3\sqrt{41}\!-\!11)\approx$0.228 \\
	         &&  & &  & &&\\
	         & $\vert 1,\frac{3}{2},\frac{3}{2}\rangle$  &B$\approx$0.304 & D$\approx$0.512 & E$\approx$0.519 & $ \frac{1}{60}(\sqrt{193}\!+\!4\sqrt{13}\!-\!15)\approx$0.222 & G$\approx$0.192 & I$\approx$0.279\\
         &&  & &  & &&\\
               & $\vert 2,\frac{3}{2},\frac{3}{2}\rangle$ & $ \frac{1}{12}(\sqrt{17}\!-\!1)\approx$0.260 &1/3$\approx$0.333 & 1/3$\approx$0.333  & 1/6$\approx$0.167&1/6$\approx$0.167&$ \frac{1}{6}(\sqrt{5}\!-\!1)\approx$0.206\\ 
             & & &  & &&\\
  & $\vert 3,\frac{3}{2},\frac{3}{2}\rangle$ & 0 & 0 & 0 & 0&0&0\\
          &&  & &  & &&\\
             \hline
                                 \multicolumn{8}{l}{$A\!=\!   \frac{1}{160}(8\sqrt{17}\!+\!\sqrt{33}\!-21)$}\\
               \multicolumn{8}{l}{$B\!=\!\tfrac{1}{135}|13\!+\!2\sqrt{247}\cos(\frac{\phi_1}{3}\!+\!\frac{2\pi}{3})|\!+\!\frac{1}{135}|13\!+\!2\sqrt{832}\cos(\frac{\phi_2}{3}\!+\!\frac{2\pi}{3})|; 
             \hspace*{2.6cm}\phi_1\!=\!\arctan\left( \frac{\sqrt{p_1^3\!-\!q_1^2}}{q_1}\right), p_1\!=\!\frac{247}{(135)^2}, q_1\!=\!\frac{3043}{(135)^3}$} \\
                   \multicolumn{8}{l}{$\hspace*{13cm} \phi_2\!=\!\arctan\left( \frac{\sqrt{p_2^3\!-\!q_2^2}}{q_2}\right), p_2\!=\!\frac{832}{(135)^2}, q_2\!=\!\frac{9388}{(135)^3}$} \\
                                   \multicolumn{8}{l}{$C\!=\!  \frac{1}{160}(3\sqrt{33}\!+\!2\sqrt{41}\!+\!5\sqrt{17}\!-\!30)\!+\! \frac{1}{240}|7\!+\!2\sqrt{240}\cos(\frac{\phi_1}{3}\!+\!\frac{2\pi}{3})|
                        ;  \hspace*{2.5cm}\phi_1\!=\!\arctan\left( \frac{\sqrt{p_1^3\!-\!q_1^2}}{q_1}\right),  p_1\!=\!\frac{247}{(240)^2}, q_1\!=\!\frac{1288}{(240)^3}$} \\
                         \multicolumn{8}{l}{$D\!=\!\frac{1}{135}|17\!+\!\sqrt{2143}\cos(\frac{\phi_1}{3}\!+\!\frac{2\pi}{3})|\!+\!\frac{1}{135}|11\!+\!\sqrt{2218}\cos(\frac{\phi_2}{3}\!+\!\frac{2\pi}{3})|\!+\! \frac{1}{90}|4\!-\!\sqrt{322}|             
                        ;\;\; \phi_1\!=\!\arctan\left( \frac{\sqrt{p_1^3\!-\!q_1^2}}{q_1}\right), p_1\!=\!\frac{2143}{(270)^2}, q_1\!=\!\frac{8614}{(270)^3}$} \\
                    \multicolumn{8}{l}{$\hspace*{13cm} \phi_2\!=\!\arctan\left( \frac{\sqrt{p_2^3\!-\!q_2^2}}{q_2}\right), p_2\!=\!\frac{1109}{2(135)^2}, q_2\!=\!\frac{5615}{(135)^3}$} \\
                         \multicolumn{8}{l}{$E\!=\!  \frac{1}{45}|4\!-\!\sqrt{217}\cos(\frac{\phi_1}{3}\!+\!\frac{6\pi}{3})|\!+\!\frac{1}{30}(\sqrt{33}\!+\!2\sqrt{19}\!-\!5)
                        ;  \hspace*{4.6cm}\phi_1\!=\!\arctan\left( \frac{\sqrt{p_1^3\!-\!q_1^2}}{q_1}\right),p_1\!=\!\frac{217}{(90)^2}, q_1\!=\!-\frac{568}{(90)^3}$} \\
                             \multicolumn{8}{l}{$F\!=\!  \frac{1}{160}(10\sqrt{2}\!+\!\sqrt{113}\!-\!17)$}\\
                         \multicolumn{8}{l}{$G\!=\!    \frac{1}{270}|31\!+\!2\sqrt{1183}\cos(\frac{\phi_1}{3}\!+\!\frac{2\pi}{3})|\!+\!\frac{1}{135}|17\!+\!\sqrt{2278}\cos(\frac{\phi_2}{3}\!+\!\frac{2\pi}{3})|
                        ;  \hspace*{2.5cm}\phi_1\!=\!\arctan\left( \frac{\sqrt{p_1^3\!-\!q_1^2}}{q_1}\right),  p_1\!=\!\frac{1183}{(270)^2}, q_1\!=\!\frac{29314}{(270)^3}$} \\
                    \multicolumn{8}{l}{$\hspace*{13.1cm} \phi_2\!=\!\arctan\left( \frac{\sqrt{p_2^3\!-\!q_2^2}}{q_2}\right), p_2\!=\!\frac{1139}{2(135)^2}, q_2\!=\!\frac{9197}{(135)^3}$} \\
                                               \multicolumn{8}{l}{$H\!=\!    \frac{1}{60}|5\!-\!\sqrt{34}\cos(\frac{\phi_1}{3}\!+\!\frac{6\pi}{3})|\!+\!\frac{1}{20}|2\!-\!\sqrt{13}|
                        ;  \hspace*{6.2cm}\phi_1\!=\!\arctan\left( \frac{\sqrt{p_1^3\!-\!q_1^2}}{q_1}\right),  p_1\!=\!\frac{34}{(120)^2}, q_1\!=\!-\frac{44}{(120)^3}$} \\
                         \multicolumn{8}{l}{$I\!=\!   \frac{1}{135}|13\!+\!2\sqrt{247}\cos(\frac{\phi_1}{3}\!+\!\frac{2\pi}{3})|\!+\!\frac{1}{270}|35\!+\!2\sqrt{1879}\cos(\frac{\phi_2}{3}\!+\!\frac{2\pi}{3})|\!+\!\frac{1}{90}|8\!-\!\sqrt{298}|
                        ;\;\;\; \phi_1\!=\!\arctan\left( \frac{\sqrt{p_1^3\!-\!q_1^2}}{q_1}\right),  p_1\!=\!\frac{247}{(135)^2}, q_1\!=\!\frac{3043}{(135)^3}$} \\
                    \multicolumn{8}{l}{$\hspace*{13.2cm} \phi_2\!=\!\arctan\left( \frac{\sqrt{p_2^3\!-\!q_2^2}}{q_2}\right), p_2\!=\!\frac{1879}{(270)^2}, q_2\!=\!\frac{42974}{(270)^3}$} \\
            
                   \hline
\end{tabular}
}
\caption{The explicit values of six   bipartite   negativities (${\cal N}_{\mu_1|S_1S_2}$, ${\cal N}_{S_1|\mu_1S_2}$, ${\cal N}_{S_2|\mu_1S_1}$, ${\cal N}_{\mu_1|\mu_2S_2}$, ${\cal N}_{\mu_2|\mu_1S_2}$  and ${\cal N}_{S_2|\mu_1\mu_2}$) calculated for all available ground states  $\vert \sigma^z_T,\sigma_1,\sigma_2\rangle$ of the mixed spin-(1/2,1) Heisenberg tetramer~\eqref{eq1}.}
\label{tab2}
\end{table*}
Moreover, it can be seen from Fig.~\ref{fig2a} that there  exists three different regimes depending on the strength of the interaction  ratio $J_1/J$. In particular, we will consider a weak interaction limit  $J_1/J\!<\!1$, a strong interaction limit $J_1/J\!>\!1$ and a fully isotropic case $J_1/J\!=\!1$. 
\begin{figure*}[t!]
{\includegraphics[width=.313\textwidth,trim=1.8cm 8.45cm 4.2cm 8cm, clip]{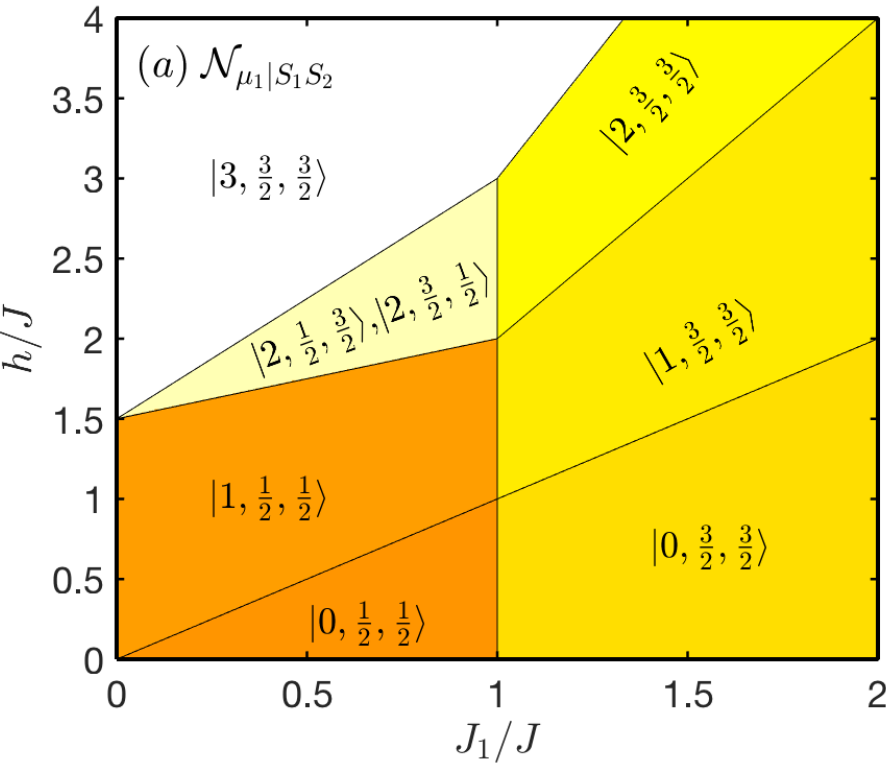}}
{\includegraphics[width=.275\textwidth,trim=3.6cm 8.45cm 4.2cm 8cm, clip]{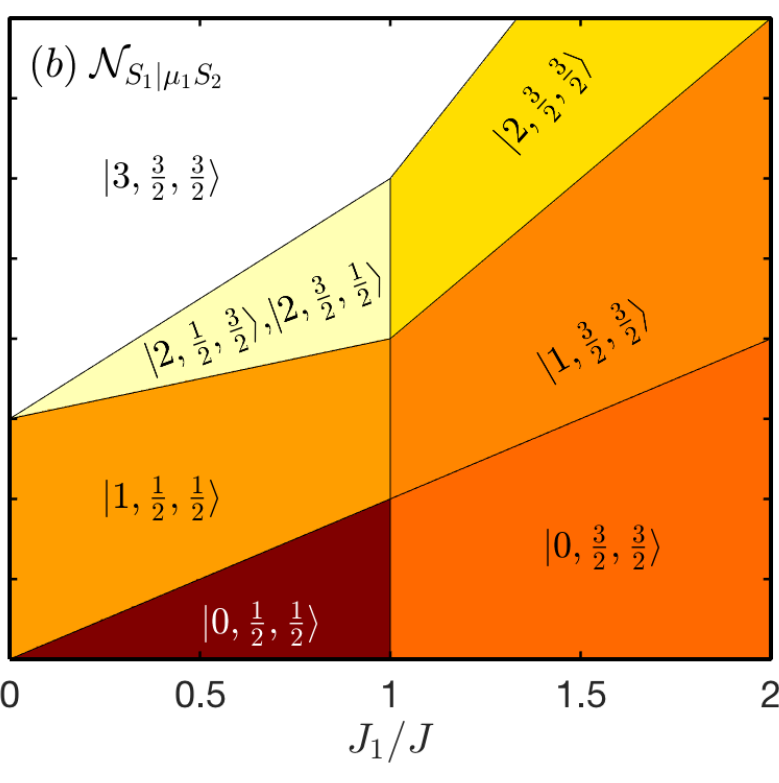}}
{\includegraphics[width=.342\textwidth,trim=3.6cm 8.45cm 1cm 8cm, clip]{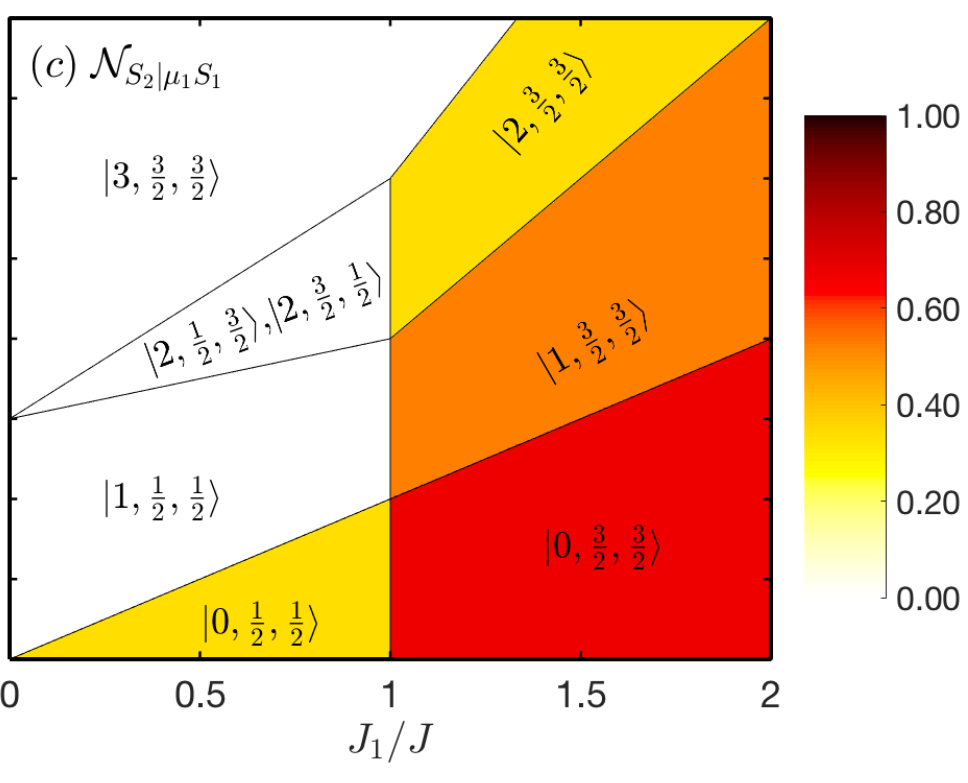}}
\\
{\includegraphics[width=.313\textwidth,trim=1.8cm 8.45cm 4.2cm 8cm, clip]{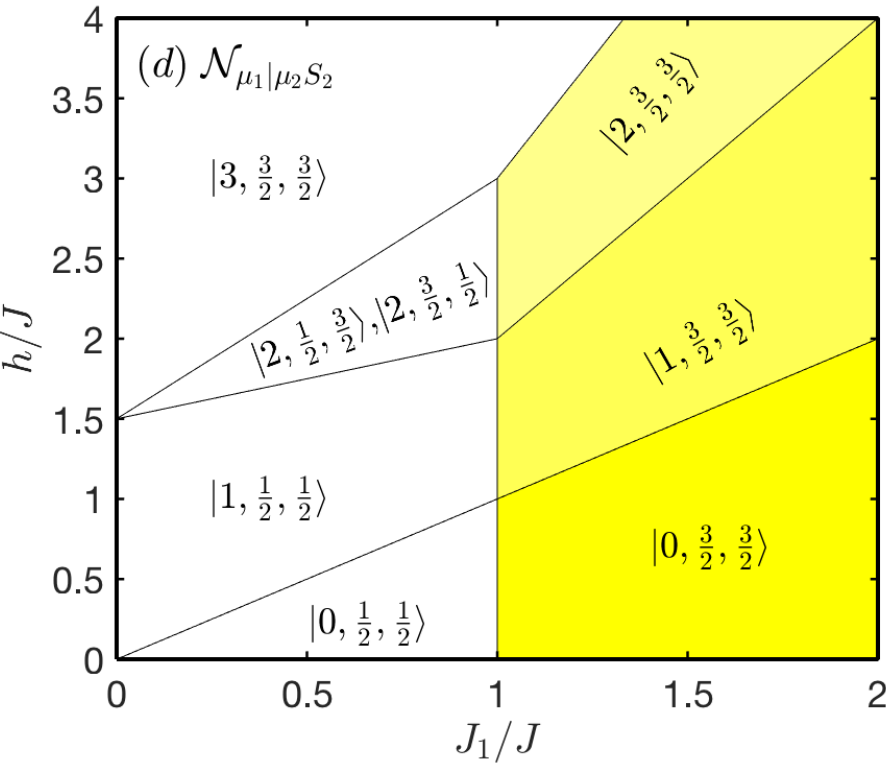}}
{\includegraphics[width=.275\textwidth,trim=3.6cm 8.45cm 4.2cm 8cm, clip]{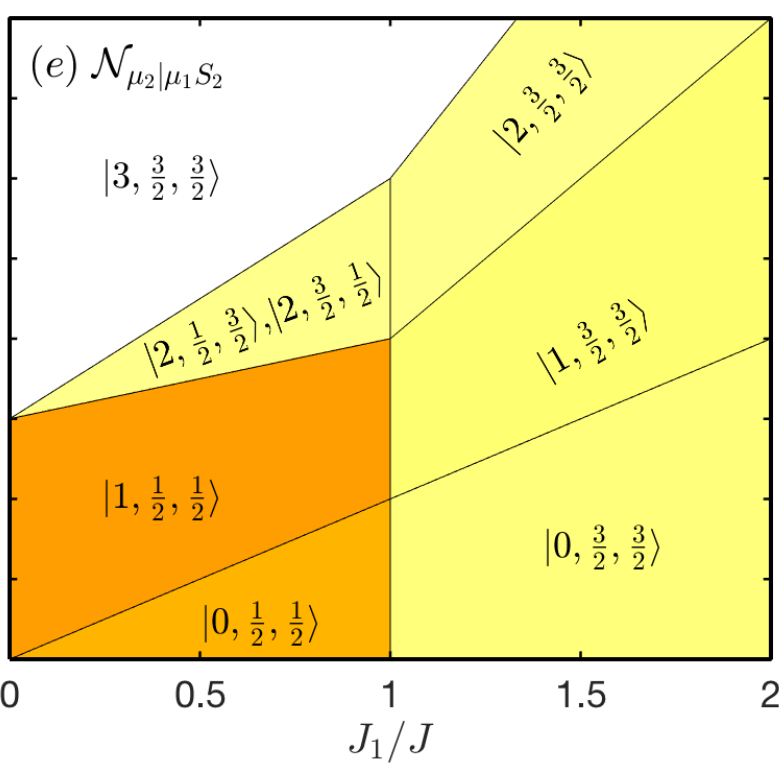}}
{\includegraphics[width=.342\textwidth,trim=3.6cm 8.45cm 1cm 8cm, clip]{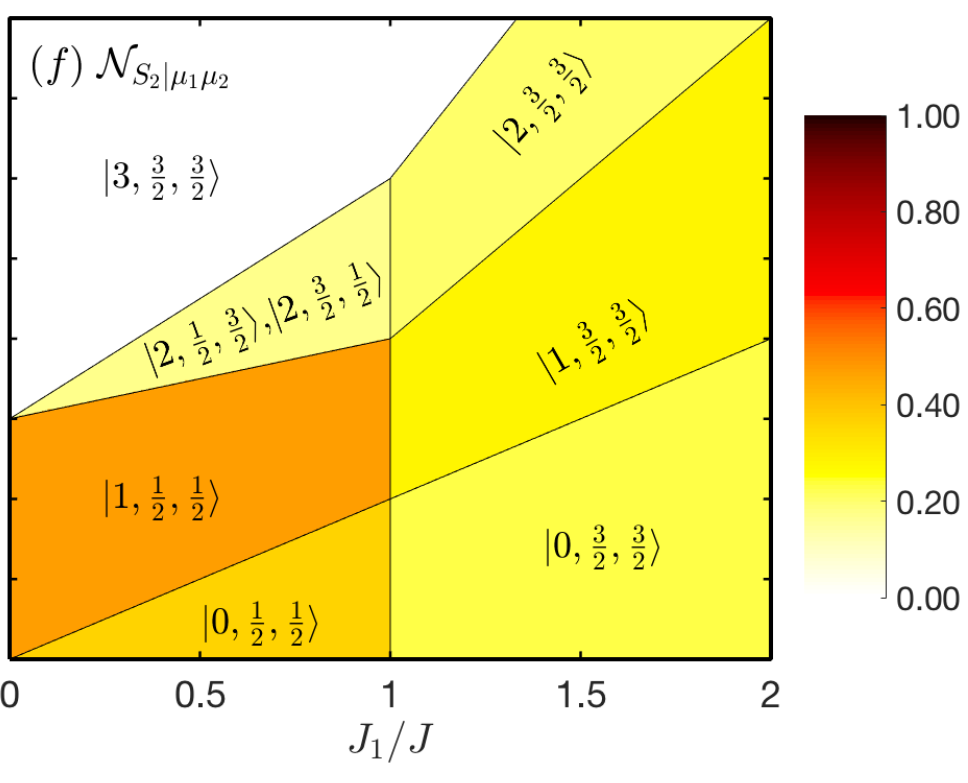}}
\caption{Density plots of    bipartite negativities ${\cal N}_{\mu_1|S_1S_2}$, ${\cal N}_{S_1|\mu_1|S_2}$ , ${\cal N}_{S_2|\mu_1S_1}$, ${\cal N}_{\mu_1|\mu_2S_2}$, ${\cal N}_{\mu_2|\mu_1S_2}$ and ${\cal N}_{S_2|\mu_1\mu_2}$   of the mixed spin-(1/2,1) Heisenberg tetramer in the $J_1/J-h/J$ plane. Solid black lines determine phase boundaries between different  ground states characterized by a set of three quantum numbers $|\sigma_T^z,\sigma_{1},\sigma_{2}\rangle$.}
\label{fig2}
\end{figure*}

In a weak interaction limit $J_1/J\!<\!1$ the mixed spin-(1/2,1) Heisenberg tetramer exhibits a nonzero genuine tripartite entanglement exclusively among three spins $\mu_1\!-\!S_1\!-\!S_2$. Under this condition,  the nonzero  value of ${\cal N}_{\mu_1S_1S_2}$ is observed  at low-enough magnetic fields where the singlet ground state $\vert 0,\frac{1}{2},\frac{1}{2}\rangle$ is realized (Fig.~\ref{fig2a}$(a)$). Within this phase, all bipartite negativities between a single spin and the remaining spin dimer ${\cal N}_{\mu_1|S_1S_2}$, ${\cal N}_{S_1|\mu_1S_2}$ and ${\cal N}_{S_2|\mu_1S_1}$ are nonzero as illustrated Fig.~\ref{fig2}$(a)-(c)$ and  the trimer  $\mu_1\!-\!S_1\!-\!S_2$ has to be fully inseparable.  From a previous analysis of the respective bipartite negativities between two  spins denoted as ${\cal N}_{A|B}$ reported in Ref.~\cite{Vargova2023}, we identify that each of  three bipartite negativities  ${\cal N}_{\mu_1|S_1S_2}$, ${\cal N}_{S_1|\mu_1S_2}$ and ${\cal N}_{S_2|\mu_1S_1}$ has at least one entangled dimer involving the  spin $\mu_1$, $S_1$ and $S_2$, respectively. Our observation has a more universal consequence and can be generalized as follows. The bipartite negativity ${\cal N}_{A|BC}$ measuring a strength of the bipartite entanglement between a single spin $A$ and the spin dimer $B\!-\!C$ of the mixed spin-(1/2,1) Heisenberg tetramer is nonzero  if at least one spin dimer $A\!-\!B$ or $A\!-\!C$ is entangled, ${\cal N}_{A|B}\!>\!0$ or ${\cal N}_{A|C}\!>\!0$. Consequently, the  genuine tripartite negativity ${\cal N}_{ABC}$ in the  mixed spin-(1/2,1) Heisenberg tetramer becomes nonzero (the tripartite system $A\!-\!B\!-\!C$ is fully inseparable) if at least two spin dimers within the tripartite system $A\!-\!B\!-\!C$ are entangled.  To support this conclusion we collect in Tab.\ref{tab1} the distribution of   bipartite negativities [positive(zero) - thick(thin) line in each small triangle]  in each trimerized spin system $A-B-C$.  The distribution of bipartite negativities is specific for each ground state  $\vert \sigma_T^z,\sigma_1,\sigma_2\rangle$. For a better lucidity the Tab.\ref{tab1} involves an additional information, namely, whether the bipartite negativity ${\cal N}_{A|BC}$ in a specific ground state $\vert \sigma_T^z,\sigma_1,\sigma_2\rangle$ is zero or not.  
\begin{table*}[t!]
\resizebox{0.98\textwidth}{!}{
\begin{tabular}{l c||l |c | l || l | c| l }
 & & \parbox[t]{2.2cm}{ground-state\\  $\vert \sigma_T^z,\sigma_1,\sigma_2\rangle$} &\parbox[t]{1.8cm}{ distribution\\ of bipartite negativities} & ${\cal N}_{A|BC}$ &  \parbox[t]{2.2cm}{ground-state\\  $\vert \sigma_T^z,\sigma_1,\sigma_2\rangle$}  &\parbox[t]{1.8cm}{ distribution\\ of bipartite negativities}  & ${\cal N}_{A|BC}$ \\
\hline
${\cal N}_{\mu_1|S_1S_2}:$ & {\multirow{4}{*}{
\begin{tikzpicture}[scale=0.5]
\draw (0,0) node[align=left,   below] {$S_1$}  -- (1,0) node[align=right,   below] {$S_2$} --(1/2,1)node[align=right,   above] {$\mu_1$} -- cycle;
\draw (1/2,1) circle [radius=0.1];
\end{tikzpicture}
} } &$\vert 2,\frac{1}{2},\frac{3}{2}\rangle$,$\vert 2,\frac{3}{2},\frac{1}{2}\rangle$ &
\begin{tikzpicture}
\draw (0,0)  -- (1/3,0)  --(1/6,1/3)-- cycle;
\draw[very thick] (0,0)  --(1/6,1/3);
\draw (1/6,1/3) circle [radius=0.07];
\end{tikzpicture}   & $\neq0$  & $\vert 2,\frac{3}{2},\frac{3}{2}\rangle$ &
{\begin{tikzpicture}
\draw[very thick]  (0,0)  -- (1/3,0)  --(1/6,1/3)-- cycle;
\draw (1/6,1/3) circle [radius=0.07];
\end{tikzpicture}}
   & $\neq0$ 
   \\
& & $\vert 1,\frac{1}{2},\frac{1}{2}\rangle$ & 
\begin{tikzpicture}
\draw (0,0)  -- (1/3,0)  --(1/6,1/3)-- cycle;
\draw[very thick] (0,0)  --(1/6,1/3);
\draw (1/6,1/3) circle [radius=0.07];
\end{tikzpicture} 
& $\neq0$  & $\vert 1,\frac{3}{2},\frac{3}{2}\rangle$ &
{\begin{tikzpicture}
\draw[very thick]  (0,0)  -- (1/3,0)  --(1/6,1/3)-- cycle;
\draw (1/6,1/3) circle [radius=0.07];
\end{tikzpicture}}  & $\neq0$ 
\\
& & $\vert 0,\frac{1}{2},\frac{1}{2}\rangle$ &
 \begin{tikzpicture}
\draw (0,0)  -- (1/3,0)  --(1/6,1/3)-- cycle;
\draw[very thick] (1/6,1/3)--(0,0)--(1/3,0);
\draw (1/6,1/3) circle [radius=0.07];
\end{tikzpicture}   
& $\neq0$  & $\vert 0,\frac{3}{2},\frac{3}{2}\rangle$ &
 \begin{tikzpicture}
\draw (0,0)  -- (1/3,0)  --(1/6,1/3)-- cycle;
\draw[very thick] (0,0)--(1/3,0)--(1/6,1/3);
\draw (1/6,1/3) circle [radius=0.07];
\end{tikzpicture} 
  & $\neq0$\\
 \hline
${\cal N}_{S_1|\mu_1S_2}$: & {\multirow{4}{*}{
\begin{tikzpicture}[scale=0.5]
\draw (0,0) node[align=left,   below] {$\mu_1$}  -- (1,0) node[align=right,   below] {$S_2$} --(1/2,1)node[align=right,   above] {$S_1$} -- cycle;
\draw (1/2,1) circle [radius=0.1];
\end{tikzpicture}
} } &$\vert 2,\frac{1}{2},\frac{3}{2}\rangle$,$\vert 2,\frac{3}{2},\frac{1}{2}\rangle$ &
\begin{tikzpicture}
\draw (0,0)  -- (1/3,0)  --(1/6,1/3)-- cycle;
\draw[very thick] (0,0)  --(1/6,1/3);
\draw (1/6,1/3) circle [radius=0.07];
\end{tikzpicture}   & $\neq0$  & $\vert 2,\frac{3}{2},\frac{3}{2}\rangle$ &
{\begin{tikzpicture}
\draw[very thick]  (0,0)  -- (1/3,0)  --(1/6,1/3)-- cycle;
\draw (1/6,1/3) circle [radius=0.07];
\end{tikzpicture}}
   & $\neq0$ 
   \\
& & $\vert 1,\frac{1}{2},\frac{1}{2}\rangle$ & 
\begin{tikzpicture}
\draw (0,0)  -- (1/3,0)  --(1/6,1/3)-- cycle;
\draw[very thick] (0,0)  --(1/6,1/3);
\draw (1/6,1/3) circle [radius=0.07];
\end{tikzpicture} 
& $\neq0$  & $\vert 1,\frac{3}{2},\frac{3}{2}\rangle$ &
{\begin{tikzpicture}
\draw[very thick]  (0,0)  -- (1/3,0)  --(1/6,1/3)-- cycle;
\draw (1/6,1/3) circle [radius=0.07];
\end{tikzpicture}}  & $\neq0$ 
\\
& & $\vert 0,\frac{1}{2},\frac{1}{2}\rangle$ &
 \begin{tikzpicture}
\draw (0,0)  -- (1/3,0)  --(1/6,1/3)-- cycle;
\draw[very thick] (0,0)--(1/6,1/3)--(1/3,0);
\draw (1/6,1/3) circle [radius=0.07];
\end{tikzpicture}   
& $\neq0$  & $\vert 0,\frac{3}{2},\frac{3}{2}\rangle$ &
 \begin{tikzpicture}
\draw (0,0)  -- (1/3,0)  --(1/6,1/3)-- cycle;
\draw[very thick] (0,0)--(1/3,0)--(1/6,1/3);
\draw (1/6,1/3) circle [radius=0.07];
\end{tikzpicture} 
  & $\neq0$\\
 \hline
${\cal N}_{S_2|\mu_1S_1}$: & {\multirow{4}{*}{
\begin{tikzpicture}[scale=0.5]
\draw (0,0) node[align=left,   below] {$\mu_1$}  -- (1,0) node[align=right,   below] {$S_1$} --(1/2,1)node[align=right,   above] {$S_2$} -- cycle;
\draw (1/2,1) circle [radius=0.1];
\end{tikzpicture}
} } &$\vert 2,\frac{1}{2},\frac{3}{2}\rangle$,$\vert 2,\frac{3}{2},\frac{1}{2}\rangle$ &
\begin{tikzpicture}
\draw (0,0)  -- (1/3,0)  --(1/6,1/3)-- cycle;
\draw[very thick] (0,0)  --(1/3,0);
\draw (1/6,1/3) circle [radius=0.07];
\end{tikzpicture}   & $=0$  & $\vert 2,\frac{3}{2},\frac{3}{2}\rangle$ &
{\begin{tikzpicture}
\draw[very thick]  (0,0)  -- (1/3,0)  --(1/6,1/3)-- cycle;
\draw (1/6,1/3) circle [radius=0.07];
\end{tikzpicture}}
   & $\neq0$ 
   \\
& & $\vert 1,\frac{1}{2},\frac{1}{2}\rangle$ & 
\begin{tikzpicture}
\draw (0,0)  -- (1/3,0)  --(1/6,1/3)-- cycle;
\draw[very thick] (0,0)  --(1/3,0);
\draw (1/6,1/3) circle [radius=0.07];
\end{tikzpicture} 
& $=0$  & $\vert 1,\frac{3}{2},\frac{3}{2}\rangle$ &
{\begin{tikzpicture}
\draw[very thick]  (0,0)  -- (1/3,0)  --(1/6,1/3)-- cycle;
\draw (1/6,1/3) circle [radius=0.07];
\end{tikzpicture}}  & $\neq0$ 
\\
& & $\vert 0,\frac{1}{2},\frac{1}{2}\rangle$ &
 \begin{tikzpicture}
\draw (0,0)  -- (1/3,0)  --(1/6,1/3)-- cycle;
\draw[very thick] (0,0)--(1/3,0)--(1/6,1/3);
\draw (1/6,1/3) circle [radius=0.07];
\end{tikzpicture}   
& $\neq0$  & $\vert 0,\frac{3}{2},\frac{3}{2}\rangle$ &
 \begin{tikzpicture}
\draw (0,0)  -- (1/3,0)  --(1/6,1/3)-- cycle;
\draw[very thick] (0,0)--(1/6,1/3)--(1/3,0);
\draw (1/6,1/3) circle [radius=0.07];
\end{tikzpicture} 
  & $\neq0$\\
 \hline
${\cal N}_{\mu_1|\mu_2S_2}$: & {\multirow{4}{*}{
\begin{tikzpicture}[scale=0.5]
\draw (0,0) node[align=left,   below] {$\mu_2$}  -- (1,0) node[align=right,   below] {$S_2$} --(1/2,1)node[align=right,   above] {$\mu_1$} -- cycle;
\draw (1/2,1) circle [radius=0.1];
\end{tikzpicture}
} } &$\vert 2,\frac{1}{2},\frac{3}{2}\rangle$,$\vert 2,\frac{3}{2},\frac{1}{2}\rangle$ &
\begin{tikzpicture}
\draw (0,0)  -- (1/3,0)  --(1/6,1/3)-- cycle;
\draw[very thick] (0,0)  --(1/3,0);
\draw (1/6,1/3) circle [radius=0.07];
\end{tikzpicture}   & $=0$  & $\vert 2,\frac{3}{2},\frac{3}{2}\rangle$ &
{\begin{tikzpicture}
\draw[very thick]  (0,0)  -- (1/3,0)  --(1/6,1/3)-- cycle;
\draw (1/6,1/3) circle [radius=0.07];
\end{tikzpicture}}
   & $\neq0$ 
   \\
& & $\vert 1,\frac{1}{2},\frac{1}{2}\rangle$ & 
\begin{tikzpicture}
\draw (0,0)  -- (1/3,0)  --(1/6,1/3)-- cycle;
\draw[very thick] (0,0)  --(1/3,0);
\draw (1/6,1/3) circle [radius=0.07];
\end{tikzpicture} 
& $=0$  & $\vert 1,\frac{3}{2},\frac{3}{2}\rangle$ &
{\begin{tikzpicture}
\draw[very thick]  (0,0)  -- (1/3,0)  --(1/6,1/3)-- cycle;
\draw (1/6,1/3) circle [radius=0.07];
\end{tikzpicture}}  & $\neq0$ 
\\
& & $\vert 0,\frac{1}{2},\frac{1}{2}\rangle$ &
 \begin{tikzpicture}
\draw (0,0)  -- (1/3,0)  --(1/6,1/3)-- cycle;
\draw[very thick] (0,0)--(1/3,0);
\draw (1/6,1/3) circle [radius=0.07];
\end{tikzpicture}   
& $=0$  & $\vert 0,\frac{3}{2},\frac{3}{2}\rangle$ &
 \begin{tikzpicture}
\draw (0,0)  -- (1/3,0)  --(1/6,1/3)-- cycle;
\draw[very thick] (0,0)--(1/6,1/3)--(1/3,0);
\draw (1/6,1/3) circle [radius=0.07];
\end{tikzpicture} 
  & $\neq0$\\
 \hline
${\cal N}_{\mu_2|\mu_1S_2}$: & {\multirow{4}{*}{
\begin{tikzpicture}[scale=0.5]
\draw (0,0) node[align=left,   below] {$\mu_1$}  -- (1,0) node[align=right,   below] {$S_2$} --(1/2,1)node[align=right,   above] {$\mu_2$} -- cycle;
\draw (1/2,1) circle [radius=0.1];
\end{tikzpicture}
} } &$\vert 2,\frac{1}{2},\frac{3}{2}\rangle$,$\vert 2,\frac{3}{2},\frac{1}{2}\rangle$ &
\begin{tikzpicture}
\draw (0,0)  -- (1/3,0)  --(1/6,1/3)-- cycle;
\draw[very thick] (1/6,1/3)  --(1/3,0);
\draw (1/6,1/3) circle [radius=0.07];
\end{tikzpicture}   & $\neq0$  & $\vert 2,\frac{3}{2},\frac{3}{2}\rangle$ &
{\begin{tikzpicture}
\draw[very thick]  (0,0)  -- (1/3,0)  --(1/6,1/3)-- cycle;
\draw (1/6,1/3) circle [radius=0.07];
\end{tikzpicture}}
   & $\neq0$ 
   \\
& & $\vert 1,\frac{1}{2},\frac{1}{2}\rangle$ & 
\begin{tikzpicture}
\draw (0,0)  -- (1/3,0)  --(1/6,1/3)-- cycle;
\draw[very thick] (1/6,1/3)  --(1/3,0);
\draw (1/6,1/3) circle [radius=0.07];
\end{tikzpicture} 
& $\neq0$  & $\vert 1,\frac{3}{2},\frac{3}{2}\rangle$ &
{\begin{tikzpicture}
\draw[very thick]  (0,0)  -- (1/3,0)  --(1/6,1/3)-- cycle;
\draw (1/6,1/3) circle [radius=0.07];
\end{tikzpicture}}  & $\neq0$ 
\\
& & $\vert 0,\frac{1}{2},\frac{1}{2}\rangle$ &
 \begin{tikzpicture}
\draw (0,0)  -- (1/3,0)  --(1/6,1/3)-- cycle;
\draw[very thick] (1/6,1/3)--(1/3,0);
\draw (1/6,1/3) circle [radius=0.07];
\end{tikzpicture}   
& $\neq0$  & $\vert 0,\frac{3}{2},\frac{3}{2}\rangle$ &
 \begin{tikzpicture}
\draw (0,0)  -- (1/3,0)  --(1/6,1/3)-- cycle;
\draw[very thick] (1/6,1/3)--(0,0)--(1/3,0);
\draw (1/6,1/3) circle [radius=0.07];
\end{tikzpicture} 
  & $\neq0$\\
 \hline
${\cal N}_{S_2|\mu_1\mu_2}$: & {\multirow{4}{*}{
\begin{tikzpicture}[scale=0.5]
\draw (0,0) node[align=left,   below] {$\mu_1$}  -- (1,0) node[align=right,   below] {$\mu_2$} --(1/2,1)node[align=right,   above] {$S_2$} -- cycle;
\draw (1/2,1) circle [radius=0.1];
\end{tikzpicture}
} } &$\vert 2,\frac{1}{2},\frac{3}{2}\rangle$,$\vert 2,\frac{3}{2},\frac{1}{2}\rangle$ &
\begin{tikzpicture}
\draw (0,0)  -- (1/3,0)  --(1/6,1/3)-- cycle;
\draw[very thick] (1/6,1/3)  --(1/3,0);
\draw (1/6,1/3) circle [radius=0.07];
\end{tikzpicture}   & $\neq0$  & $\vert 2,\frac{3}{2},\frac{3}{2}\rangle$ &
{\begin{tikzpicture}
\draw[very thick]  (0,0)  -- (1/3,0)  --(1/6,1/3)-- cycle;
\draw (1/6,1/3) circle [radius=0.07];
\end{tikzpicture}}
   & $\neq0$ 
   \\
& & $\vert 1,\frac{1}{2},\frac{1}{2}\rangle$ & 
\begin{tikzpicture}
\draw (0,0)  -- (1/3,0)  --(1/6,1/3)-- cycle;
\draw[very thick] (1/6,1/3)  --(1/3,0);
\draw (1/6,1/3) circle [radius=0.07];
\end{tikzpicture} 
& $\neq0$  & $\vert 1,\frac{3}{2},\frac{3}{2}\rangle$ &
{\begin{tikzpicture}
\draw[very thick]  (0,0)  -- (1/3,0)  --(1/6,1/3)-- cycle;
\draw (1/6,1/3) circle [radius=0.07];
\end{tikzpicture}}  & $\neq0$ 
\\
& & $\vert 0,\frac{1}{2},\frac{1}{2}\rangle$ &
 \begin{tikzpicture}
\draw (0,0)  -- (1/3,0)  --(1/6,1/3)-- cycle;
\draw[very thick] (1/6,1/3)--(1/3,0);
\draw (1/6,1/3) circle [radius=0.07];
\end{tikzpicture}   
& $\neq0$  & $\vert 0,\frac{3}{2},\frac{3}{2}\rangle$ &
 \begin{tikzpicture}
\draw (0,0)  -- (1/3,0)  --(1/6,1/3)-- cycle;
\draw[very thick] (1/6,1/3)--(0,0)--(1/3,0);
\draw (1/6,1/3) circle [radius=0.07];
\end{tikzpicture} 
  & $\neq0$\\
 \hline
\end{tabular}
}
\caption{The graphical visualization of distribution  of the bipartite negativities ${\cal N}_{A|B}$ determining the  bipartite negativity ${\cal N}_{A|BC}$ between the spin $A$ and the spin dimer $B\!-\!C$ of the mixed spin-(1/2,1) Heisenberg tetramer~\eqref{eq1}. The specific  distribution of the bipartite negativities ${\cal N}_{A|B}$ is given for the respective ground state $\vert \sigma_T^z,\sigma_1,\sigma_2\rangle$. The nonzero (zero) value of the  bipartite negativity ${\cal N}_{A|B}$ is visualized through the thick (thin) edge of a triangle (column No. 3 and No. 6). The spin species $A$ of the bipartite negativity   ${\cal N}_{A|BC}$  is highlighted by a circle at a top of each triangle. The information about the zero (nonzero) value of  the bipartite negativity   ${\cal N}_{A|BC}$ in a specific ground state is also inserted in a table (column No. 4 and No. 7).}
\label{tab1}
\end{table*}
\\

In the strong interaction limit $J_1/J\!>\!1$  the nonzero genuine tripartite negativity of the mixed spin-(1/2,1) Heisenberg tetramer is detected for each $h/J\!\leq\!3J_1/J$ in both tripartite negativities ${\cal N}_{\mu_1S_1S_2}$ and ${\cal N}_{\mu_1\mu_2S_2}$, respectively.  Fig.~\ref{fig2} demonstrates that the nonzero genuine tripartite negativity at  $J_1/J\!>\!1$ arises from the positive value of all three  bipartite negativities ${\cal N}_{A|BC}$, ${\cal N}_{B|AC}$ and ${\cal N}_{C|AB}$ in the spin subsystem $\mu_1\!-\!S_1\!-\!S_2$ (Fig.~\ref{fig2}$(a)$-$(c)$) as well as the spin subsystem $\mu_1\!-\!\mu_2\!-\!S_2$ (Fig.~\ref{fig2}$(d)$-$(f)$). 
Similarly as in the weak interaction limit, the nonzero value  of the bipartite negativity ${\cal N}_{A|BC}$ is again determined by one or two nonzero bipartite negativities ${\cal N}_{A|B}$ or ${\cal N}_{A|C}$ (see the last three columns of Tab.~\ref{tab1}). Although the distribution of the respective bipartite negativities  ${\cal N}_{A|B}$ and ${\cal N}_{A|C}$ can be used to predict the nonzero  bipartite negativity ${\cal N}_{A|BC}$ between the single spin $A$ and the spin dimer $B\!-\!C$ it should be emphasized that in case of a quantitative estimation one should be very careful, because identical values of the bipartite negativities ${\cal N}_{A|B}$ and ${\cal N}_{A|C}$ can determine various bipartite negativities ${\cal N}_{A|BC}$. For example,  one finds
\begin{align}
&{\cal N}_{\mu_1|S_1S_2}\left(\left\vert1,\frac{1}{2},\frac{3}{2}\right \rangle,\left\vert1,\frac{3}{2},\frac{1}{2} \right\rangle\right)\!=\!\frac{1}{6}(\sqrt{3}\!-\!1)
\nonumber\\
& \;\;\rightarrow\;\;{\cal N}_{\mu_1|S_1}\left(\left\vert1,\frac{1}{2},\frac{3}{2}\right \rangle,\left\vert1,\frac{3}{2},\frac{1}{2} \right\rangle\right)\!=\!\frac{1}{12}(\sqrt{17}\!-\!3),  
\nonumber\\
 &\;\;\;\;\;\;\;\;\;\;{\cal N}_{\mu_1|S_2}\left(\left\vert1,\frac{1}{2},\frac{3}{2}\right \rangle,\left\vert1,\frac{3}{2},\frac{1}{2} \right\rangle\right)\!=\!0.
 \label{eq6}
 \end{align}
 \begin{align}
&{\cal N}_{S_2|\mu_1\mu_2}\left(\left\vert1,\frac{1}{2},\frac{3}{2}\right \rangle,\left\vert1,\frac{3}{2},\frac{1}{2} \right\rangle\right)\!=\!\frac{1}{6} 
\nonumber\\
&\;\;\rightarrow\;\;{\cal N}_{\mu_1|S_1}\left(\left\vert1,\frac{1}{2},\frac{3}{2}\right \rangle,\left\vert1,\frac{3}{2},\frac{1}{2} \right\rangle\right)\!=\!\frac{1}{12}(\sqrt{17}\!-\!3),  
\nonumber\\
 &\;\;\;\;\;\;\;\;\;\;{\cal N}_{\mu_1|S_2}\left(\left\vert1,\frac{1}{2},\frac{3}{2}\right \rangle,\left\vert1,\frac{3}{2},\frac{1}{2} \right\rangle\right)\!=\!0.
 \label{eq7}
 \end{align}

It should be nevertheless pointed out that  the  genuine tripartite negativity ${\cal N}_{\mu_1S_1S_2}$ of the mixed spin-(1/2,1) Heisenberg tetramer is suppressed with an increasing magnetic field, which additionally causes  discontinuous changes at the relevant transition fields. In contrast,  the  genuine tripartite negativity ${\cal N}_{\mu_1\mu_2S_2}$ of the mixed spin-(1/2,1) Heisenberg tetramer saturates at finite magnetic field, where the triplet ground state $\vert 1,\frac{3}{2},\frac{3}{2}\rangle$ is favored. Of course, this non-trivial behavior reflects the behavior of respective subparts ${\cal N}_{\mu_1|\mu_2S_2}$ and ${\cal N}_{S_2|\mu_1\mu_2}$, in which the bipartite negativities ${\cal N}_{\mu_1|\mu_2}$, ${\cal N}_{\mu_1|S_1}$ and ${\cal N}_{\mu_1|S_2}$ determine the final properties. As was discussed in Ref.~\cite{Vargova2023}, the increasing magnetic field enhances the bipartite negativity ${\cal N}_{\mu_1|S_1}$ and ${\cal N}_{\mu_1|S_2}$, whereas the completely opposite tendency is observed for ${\cal N}_{\mu_1|\mu_2}$. Consequently, the competition between these two effects results in an unexpected enhancement of the genuine tripartite negativity ${\cal N}_{\mu_1\mu_2S_2}$ under the  external magnetic field.

For  completeness, we briefly examine the special  fully isotropic case  $J_1/J\!=\!1$. In this specific limit, the ground state  is always degenerate for sufficiently low magnetic fields  $h/J\!<\!3$, whereby the respective degeneracy causes  reduction of the degree of  genuine tripartite negativities  ${\cal N}_{\mu_1S_1S_2}$ and ${\cal N}_{\mu_1\mu_2S_2}$ with respect to the anisotropic case $J_1/J\!\neq\!1$. In an external magnetic field, the  genuine tripartite negativities ${\cal N}_{\mu_1S_1S_2}$ and ${\cal N}_{\mu_1\mu_2S_2}$  change discontinuously at the relevant magnetic-field-driven phase transitions, whereby the rising magnetic field reduces in general a degree of both  genuine tripartite negativities. Due to a smaller total quantum spin number of the respective tripartite system,  the   genuine tripartite negativity ${\cal N}_{\mu_1\mu_2S_2}$  achieves  smaller values than the one ${\cal N}_{\mu_1S_1S_2}$ over the whole parametric space.
\\

\begin{figure*}[t!]
{\includegraphics[width=.31\textwidth,trim=1.9cm 8.45cm 3.9cm 8cm, clip]{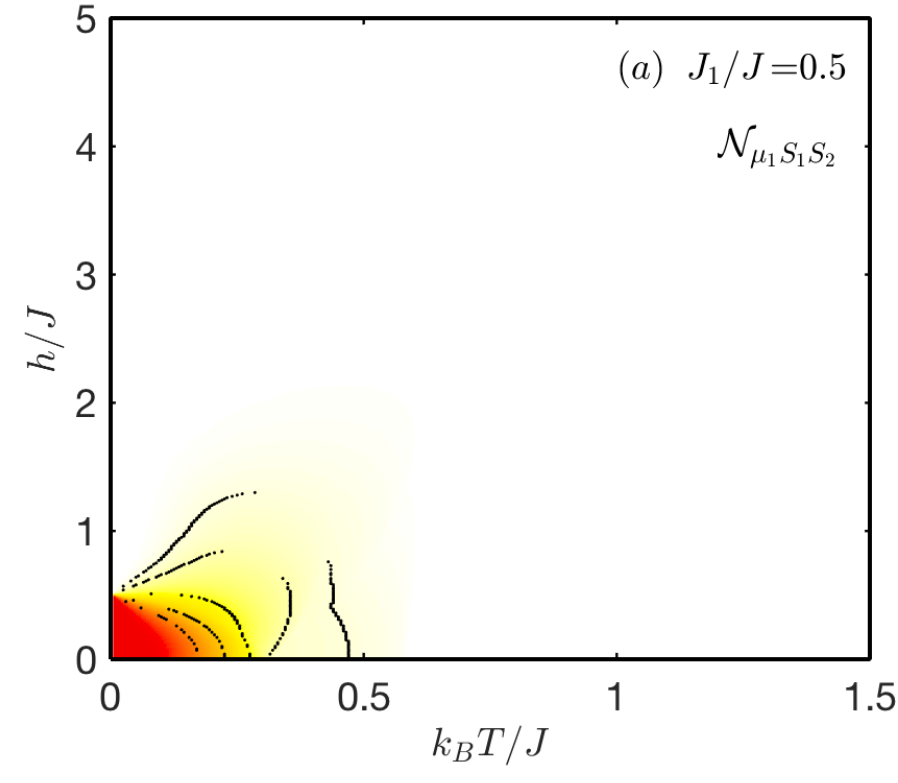}}
{\includegraphics[width=.29\textwidth,trim=3cm 8.45cm 3.9cm 8cm, clip]{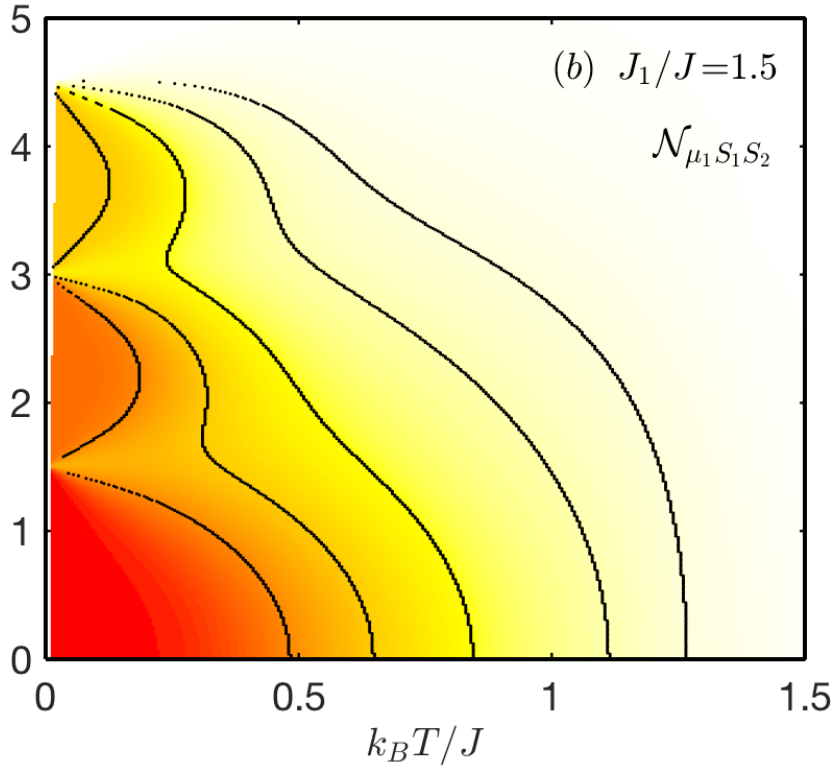}}
{\includegraphics[width=.35\textwidth,trim=3cm 8.45cm 1cm 8cm, clip]{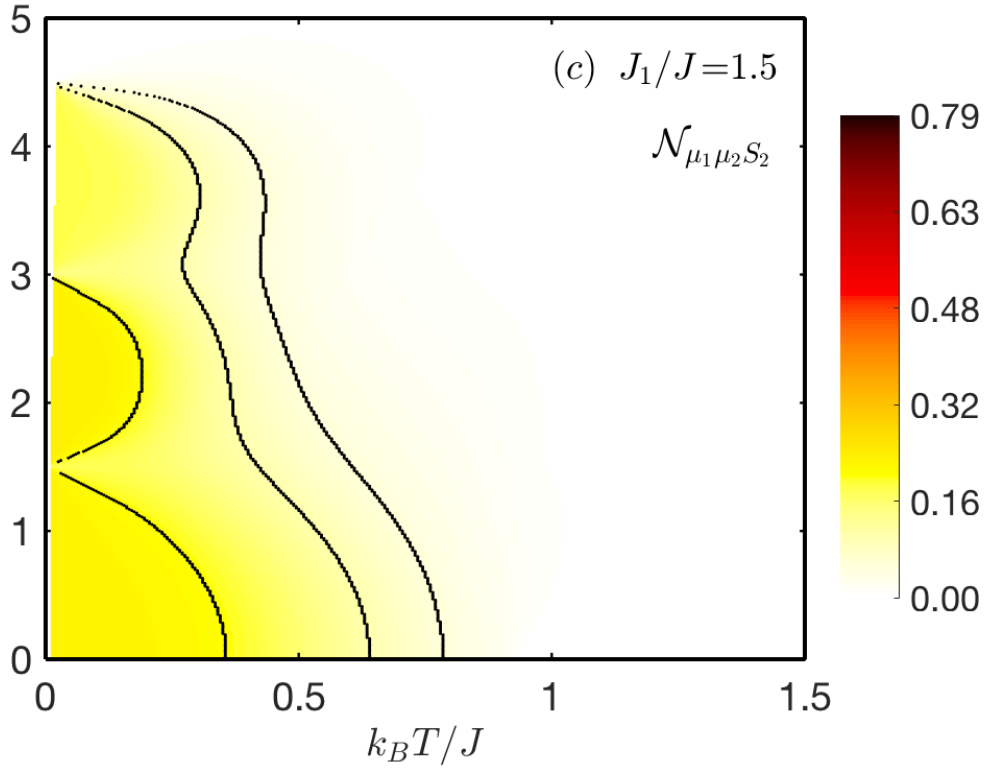}}
\caption{Density plots of genuine tripartite negativities ${\cal N}_{\mu_1S_1S_2}$ and ${\cal N}_{\mu_1\mu_2S_2}$ of the mixed spin-(1/2,1) Heisenberg tetramer~\eqref{eq1} in the $k_BT/J-h/J$ plane for two selected values of the interaction ratio $J_1/J\!=\!0.5$ and $J_1/J\!=\!1.5$. The black contour  lines correspond to values 0.05, 0.1, 0.2, 0.3 and 0.4 (from left to right).}
\label{fig3}
\end{figure*}

Now, let us look at a possibility to stabilize the genuine tripartite negativity of the mixed spin-(1/2,1) Heisenberg tetramer at  nonzero temperatures. The density plots of  thermal genuine tripartite negativities ${\cal N}_{\mu_1S_1S_2}$ and ${\cal N}_{\mu_1\mu_2S_2}$ are shown in Fig.~\ref{fig3} for two specific values of the interaction ratio $J_1/J\!=\!0.5$ and $J_1/J\!=\!1.5$, respectively. It should be emphasized that in the weak interaction limit $J_1/J\!<\!1$ the  genuine tripartite negativity is nonzero only for the three spins $\mu_1\!-\!S_1\!-\!S_2$ within the $\vert0,\frac{1}{2},\frac{1}{2}\rangle$ ground state. 
It follows from the  density plot displayed in Fig.~\ref{fig3}$(a)$ for the genuine tripartite negativity   ${\cal N}_{\mu_1S_1S_2}$ that  the  thermal genuine tripartite entanglement in the mixed spin-(1/2,1)  Heisenberg tetramer   can be detected for sufficiently low values of the interaction ratio $J_1/J$ only for relatively small values of the magnetic field and temperature $h/J\!\lesssim\!J_1/J$ and $k_BT/J\!\lesssim\!J_1/J$. In addition, the increasing temperature gradually  reduces the degree of the genuine tripartite negativity ${\cal N}_{\mu_1S_1S_2}$ until the threshold  temperature is reached, above which the trimerized spin subsystem $\mu_1\!-\!S_1\!-\!S_2$ becomes fully separable.  It should be nevertheless pointed out that the genuine tripartite negativity  ${\cal N}_{\mu_1S_1S_2}$ exhibits an intriguing thermally driven reentrant behavior slightly above the threshold magnetic field. In this parameter region, the  increasing thermal fluctuations populate  low-lying eigenstates, which cause unconventional enhancement of the tripartite entanglement above the ground state with a fully separable character. The aforementioned nontrivial thermal  behavior is clearly demonstrated through the  contour lines in Fig.~\ref{fig3}$(a)$ corresponding to  small enough values of the genuine tripartite negativity. 
It also follows from Fig.~\ref{fig3}  that the genuine tripartite negativities are thermally more resistant for the opposite interaction limit with a sufficiently high value of the interaction ratio $J_1/J\!>\!1$. In both measures of the genuine tripartite thermal entanglement ${\cal N}_{\mu_1S_1S_2}$ (Fig.~\ref{fig3}$(b)$)   and ${\cal N}_{\mu_1\mu_2S_2}$ (Fig.~\ref{fig3}$(c)$), the threshold temperatures tend to significantly higher values in comparison to the previous case $J_1/J\!<\!1$. It is quite evident that  the threshold temperature of the genuine tripartite negativity ${\cal N}_{\mu_1S_1S_2}$  is almost two times higher than for ${\cal N}_{\mu_1\mu_2S_2}$. Moreover, the strength of the thermal genuine negativity  is likewise higher for ${\cal N}_{\mu_1S_1S_2}$, which reflects the magnitude of a  genuine tripartite entanglement. A possible explanation of this difference may lie in a different total quantum spin number of the respective spin trimers $\mu_1\!-\!S_1\!-\!S_2$ and $\mu_1\!-\!\mu_2\!-\!S_2$, where  the contribution of the bipartite entanglement between two spins-1 (the first case) and two spins-1/2 (the former case) represents the crucial factor determining a stability of the genuine tripartite entanglement. 

\section{Conclusion}
\label{conclusion}
It the present paper we have rigorously examined the presence of the genuine tripartite entanglement in the mixed spin-(1/2,1) Heisenberg tetramer. The genuine tripartite entanglement among three spins $A\!-\!B\!-\!C$ has been classified through the  genuine tripartite negativity ${\cal N}_{ABC}$ defined as a  geometric mean of three possible bipartite negativities between a single spin  and the respective spin dimer, i.e.  ${\cal N}_{A|BC}$, ${\cal N}_{B|AC}$  and ${\cal N}_{C|AB}$. The bipartite negativity ${\cal N}_{A|BC}$ between a single spin $A$ and a spin dimer $B\!-\!C$ has been validated through the quantification criterion by Vidal and Werner, from which the sum of absolute values of all negative eigenvalues of a partially transposed density matrix (say $\hat{\rho}_{A|BC}^{T_A}$) unambiguously defines the degree of  a bipartite negativity (say ${\cal N}_{A|BC}$). It is convenient to perform the partial transposition with respect to the selected  spin $A$ as it is noticed by an upper superscript $T_A$. Due to the symmetry of the square plaquette shown in Fig.~\ref{fig1}, the tripartite entanglement of the mixed spin-(1/2,1) Heisenberg tetramer is characterized by two different genuine tripartite negativities ${\cal N}_{\mu_1S_1S_2}$ and ${\cal N}_{\mu_1\mu_2S_2}$. The zero-temperature and nonzero temperature behavior of both genuine tripartite  negativities has been analyzed in detail with respect to the strength of the interaction ratio $J_1/J$ and the magnetic field $h/J$. It should be recalled that the coupling constants $J$ and $J_1$  denote the isotropic  exchange interactions between the nearest-neighbor spins with different magnitudes from the same dimers and the further-neighbor spins from different dimers, respectively. 
It turns out that  in the weak interaction limit ($J_1/J\!<\!1$) the  genuine tripartite entanglement can exist solely within the spin trimer $\mu_1\!-\!S_1\!-\!S_2$ for   low-enough magnetic fields, where the singlet ground state $\vert 0,\frac{1}{2},\frac{1}{2}\rangle$ is realized. It was demonstrated that the nonzero value of the tripartite negativity ${\cal N}_{\mu_1S_1S_2}$ arises from nonzero value of all possible  bipartite negativities ${\cal N}_{\mu_1|S_1S_2}$, ${\cal N}_{S_1|\mu_1S_2}$ and ${\cal N}_{S_2|\mu_1S_1}$. In contrast to this, in the strong interaction limit ($J_1/J\!>\!1$) the  genuine tripartite entanglement can be found in both nonequivalent spin trimers $\mu_1\!-\!S_1\!-\!S_2$  and $\mu_1\!-\!\mu_2\!-\!S_2$, respectively. It has been also proved that the  genuine tripartite entanglement exists in both inequivalent spin trimers also in the fully isotropic case $J_1/J\!=\!1$ with the unique coupling constant.  Nevertheless, it is worthwhile to remark that the size  of the genuine tripartite negativity is significantly reduced in comparison to the aforementioned cases due to the respective ground-state degeneracy. 

Finally, an investigation of a thermal stability of the genuine tripartite entanglement in the mixed spin-(1/2,1) Heisenberg tetramer resulted in two interesting findings. First, the threshold temperature displays a gradual rise upon strengthening  of the interaction ratio  $J_1/J$. Second, the higher total quantum spin number of the respective spin trimer has an additional enhancement effect on a magnitude of  the  threshold temperature. The strength of the genuine tripartite entanglement is also enhanced in the spin trimer with a higher value of the total quantum spin number. Another interesting finding is that  in the weak interaction limit $J_1/J\!<\!1$  thermal fluctuations populating  low-lying eigenstates may cause the nonzero genuine tripartite negativity above the saturation magnetic field, where the  genuine tripartite entanglement is absent at zero temperature.  
\section*{Acknowledgments}
This work was financially supported by the grant of the Slovak Research and Development Agency provided under the contract No. APVV-20-0150 and by the grant of The Ministry of Education, Science, Research, and Sport of the Slovak Republic provided under the contract No. VEGA 1/0105/20.

\appendix
\section{Bipartite entanglement ${\cal N}_{\mu_{1}|\mu_{2}S_{2}}$ between the single spin $\mu_1$ and the spin dimer $\mu_2-S_2$}
\label{App A}
\setcounter{equation}{0}
\renewcommand{\theequation}{\thesection.\arabic{equation}}
\renewcommand{\thetable}{\thesection.\arabic{equation}}

The reduced density operator $\hat{\rho}_{\mu_{1}|\mu_{2}S_{2}}$ is defined after tracing out degree of freedom of the spin $S_{1}$. Thus,
\begin{align}
\hat{\rho}_{\mu_{1}|\mu_{2}S_{2}}\!&=\!\sum_{S_{1}^z} \langle S_{1}^z\vert\hat{\rho}\vert S_{1}^z\rangle
\nonumber\\
\!&=\!\frac{1}{\cal Z}\sum_{k=1}^{36}  {\rm e}^{-\beta \varepsilon_k}\left(\sum_{S_1^z}
\langle S_1^z|\psi_k\rangle\langle\psi_k|S_1^z\rangle\right).
\label{a1}
\end{align}
The  corresponding reduced density matrix $\hat{\rho}_{\mu_{1}|\mu_{2}S_{2}}$  in a basis of $\vert \mu^z_{1},\mu_{2}^z,S^z_{2}\rangle$ reads as follows
\begin{widetext}
\begin{align}
\allowdisplaybreaks
{\hat{\rho}_{\mu_{1}|\mu_{2},S_{2}}}\!=\!
\resizebox{0.9\textwidth}{!}{$
\begin{blockarray}{r ccc ccc ccc ccc}
 & \vert \frac{1}{2},\frac{1}{2},1\rangle & \vert \frac{1}{2},\frac{1}{2},0\rangle &\vert \frac{1}{2},\frac{1}{2},\!-\!1\rangle  
 & \vert \frac{1}{2},\!-\!\frac{1}{2},1\rangle & \vert \frac{1}{2},\!-\!\frac{1}{2},0\rangle &\vert \frac{1}{2},\!-\!\frac{1}{2},\!-\!1\rangle 
 &\vert \!-\!\frac{1}{2},\frac{1}{2},1\rangle&\vert \!-\!\frac{1}{2},\frac{1}{2},0\rangle&\vert \!-\!\frac{1}{2},\frac{1}{2},\!-\!1\rangle
  &\vert \!-\!\frac{1}{2},\!-\!\frac{1}{2},1\rangle&\vert \!-\!\frac{1}{2},\!-\!\frac{1}{2},0\rangle&\vert \!-\!\frac{1}{2},\!-\!\frac{1}{2},\!-\!1\rangle\\
\begin{block}{r(ccc ccc ccc ccc)}
\langle \frac{1}{2}, \frac{1}{2},1\vert \;\;\;&\rho_{1,1} & 0 & 0 & 0 & 0 & 0& 0 & 0 & 0& 0 & 0 & 0 \\
\langle \frac{1}{2}, \frac{1}{2},0\vert \;\;\;&0 & \rho_{2,2} & 0 & \rho_{2,4} & 0 & 0 & \rho_{2,7} & 0 & 0 & 0 & 0 & 0\\
\langle \frac{1}{2}, \frac{1}{2},\!-\!1\vert\;\;\; & 0 & 0 & \rho_{3,3} &  0 &\rho_{3,5}& 0&  0 &\rho_{3,8}& 0& \rho_{3,10} & 0 & 0 \\
 \langle \frac{1}{2},\!-\!\frac{1}{2},1\vert\; \;\;& 0& \rho_{4,2} &0 & \rho_{4,4} & 0 & 0 &\rho_{4,7} & 0 & 0 & 0 & 0 & 0\\
\langle \frac{1}{2},\!-\!\frac{1}{2},0\vert\;\;\;&0 &0 & \rho_{5,3} &0  &  \rho_{5,5} & 0 &  0 &\rho_{5,8}& 0& \rho_{5,10} & 0 & 0\\
 \langle \frac{1}{2},\!-\!\frac{1}{2},\!-\!1\vert \;\;\;& 0 & 0 & 0 & 0 & 0 & \rho_{6,6}& 0 & 0 & \rho_{6,9} & 0 & \rho_{6,11}& 0 \\
 \langle \!-\!\frac{1}{2}, \frac{1}{2},1\vert \;\;\;&0 & \rho_{7,2} & 0 & \rho_{7,4} & 0 & 0 & \rho_{7,7} & 0 & 0 & 0 & 0 & 0
 \\
  \langle \!-\!\frac{1}{2}, \frac{1}{2},0\vert \;\;\;& 0 & 0 & \rho_{8,3} &  0 &\rho_{8,5}& 0&  0 &\rho_{8,8}& 0& \rho_{8,10} & 0 & 0
 \\
  \langle \!-\!\frac{1}{2}, \frac{1}{2},\!-\!1\vert \;\;\;&0 & 0 & 0 & 0 & 0 & \rho_{9,6}& 0 & 0 & \rho_{9,9} & 0 & \rho_{9,11}& 0
 \\
  \langle \!-\!\frac{1}{2}, \!-\!\frac{1}{2},1\vert \;\;\;& 0 & 0 & \rho_{10,3} &  0 &\rho_{10,5}& 0&  0 &\rho_{10,8}& 0& \rho_{10,10} & 0 & 0
 \\
  \langle \!-\!\frac{1}{2}, \!-\!\frac{1}{2},0\vert \;\;\;&0 & 0 & 0 & 0 & 0 & \rho_{11,6}& 0 & 0 & \rho_{11,9} & 0 & \rho_{11,11}& 0
 \\
  \langle \!-\!\frac{1}{2}, \!-\!\frac{1}{2},\!-\!1\vert \;\;\;&0 & 0 & 0 & 0 & 0& 0 & 0 & 0& 0 & 0 & 0 & \rho_{12,12}
 \\
\end{block}
\end{blockarray}\;\;,$}
\label{a2}
\end{align}
where
\begin{flalign}
\rho_{1,1}\!&=\!\frac{{\rm e}^{ \frac{\beta }{4}J_1}}{30{\cal Z}}\bigg\{
\cosh\left(\frac{\beta h}{2}\right){\rm e}^{\frac{5}{2}\beta h}\left[60{\rm e}^{-\beta(J+\frac{5}{2}J_1)} \right]
\nonumber\\
\!&+\!\cosh\left(\frac{\beta h}{2}\right){\rm e}^{\frac{3}{2}\beta h}\left[15{\rm e}^{-\beta(J-\frac{J_1}{2})}\!-\!18{\rm e}^{-\beta(J+\frac{5}{2}J_1)}\!+\!15{\rm e}^{-\beta(-\frac{J}{2}+J_1)} \right]
\nonumber\\
\!&+\!\sinh\left(\frac{\beta h}{2}\right){\rm e}^{\frac{3}{2}\beta h}\left[5{\rm e}^{-\beta(J-\frac{J_1}{2})}\!-\!22{\rm e}^{-\beta(J+\frac{5}{2}J_1)}\!+\! 5{\rm e}^{-\beta(-\frac{J}{2}+J_1)}\right]
\nonumber\\
\!&+\!6{\rm e}^{-\frac{\beta}{4}(J-7J_1)}{\rm e}^{\beta h}\left[ 
3\cosh\left(\frac{3\beta}{4}(J\!-\!J_1)\right)\right.
\!+\!\left.2\sinh\left(\frac{3\beta}{4}(J\!-\!J_1)\right)\right]\bigg\},
\label{a3}\\
\rho_{12,12}\!&=\!\rho_{1,1}(-h),
\label{a4}\\
\rho_{2,2}\!&=\!\frac{{\rm e}^{ \frac{\beta }{4}J_1}}{90{\cal Z}}\bigg\{
\cosh\left(\frac{\beta h}{2}\right){\rm e}^{\frac{3}{2}\beta h}\left[25{\rm e}^{-\beta(J-\frac{J_1}{2})}\!+\!45{\rm e}^{-\beta(J+\frac{5}{2}J_1)}\!+\!45{\rm e}^{-\beta(-\frac{J}{2}+J_1)} \right]
\nonumber\\
\!&+\!\cosh\left(\frac{\beta h}{2}\right){\rm e}^{\frac{\beta}{2} h}\left[20{\rm e}^{-\beta(-2J+\frac{J_1}{2})}\!+\!10{\rm e}^{-\beta(J-\frac{J_1}{2})} \!+\!18{\rm e}^{-\beta(J_x+\frac{5}{2}J_1)}\!+\!17{\rm e}^{-\beta(J-\frac{5}{2}J_1)}\!+\!40{\rm e}^{-\beta(-\frac{J}{2}+J_1)}\right.
\nonumber\\
&\hspace*{3cm}\left.\!+\!35{\rm e}^{-\beta(-\frac{J}{2}-J_1)}\right]
\nonumber\\
\!&+\!\sinh\left(\frac{\beta h}{2}\right){\rm e}^{\frac{3}{2}\beta h}\left[35{\rm e}^{-\beta(J-\frac{J_1}{2})}\!+\!15{\rm e}^{-\beta(J+\frac{5}{2}J_1)}\!+\!15{\rm e}^{-\beta(-\frac{J}{2}+J_1)} \right]
\nonumber\\
\!&+\!\sinh\left(\frac{\beta h}{2}\right){\rm e}^{\frac{\beta}{2} h}\left[15{\rm e}^{-\beta(J-\frac{5}{2}J_1)}\!-\!10{\rm e}^{-\beta(-\frac{J}{2}+J_1)}\!-\!15{\rm e}^{-\beta(-\frac{J}{2}-J_1)}\right]
\nonumber\\
\!&+\!5{\rm e}^{-\beta(-\frac{J}{2}-2J_1)}\left[
3\cosh\left(\frac{3\beta}{2}(J\!-\!J_1)\right)\!+\!\sinh\left(\frac{3\beta}{2}(J\!-\!J_1)\right)
\right]
\bigg\},
\label{a5}\\
\rho_{11,11}\!&=\!\rho_{2,2}(-h),
\label{a6}\\
\rho_{3,3}\!&=\!\frac{{\rm e}^{ \frac{\beta }{4}J_1}}{90{\cal Z}}\bigg\{
\cosh\left(\frac{\beta h}{2}\right){\rm e}^{\frac{\beta }{2}h}\left[20{\rm e}^{-\beta(J-\frac{J_1}{2})}\!+\!9{\rm e}^{-\beta(J+\frac{5}{2}J_1)}\!+\!6{\rm e}^{-\beta(J-\frac{5}{2}J_1)}\!+\!30{\rm e}^{-\beta(-\frac{J}{2}+J_1)}\!+\!80{\rm e}^{-\beta(-\frac{J}{2}-J_1)} \right]
\nonumber\\
\!&+\!\cosh\left(\frac{\beta h}{2}\right){\rm e}^{-\frac{\beta }{2}h}\left[50{\rm e}^{-\beta(-2J+\frac{J_1}{2})}\!+\!12{\rm e}^{-\beta(J+\frac{5}{2}J_1)}\!+\!8{\rm e}^{-\beta(J-\frac{5}{2}J_1)}\!+\!40{\rm e}^{-\beta(-\frac{J}{2}+J_1)} \right]
\nonumber\\
\!&+\!\sinh\left(\frac{\beta h}{2}\right){\rm e}^{\frac{\beta }{2}h}\left[10{\rm e}^{-\beta(J-\frac{J_1}{2})}\!+\!3{\rm e}^{-\beta(J+\frac{5}{2}J_1)}\!+\!12{\rm e}^{-\beta(J-\frac{5}{2}J_1)}\!+\!10{\rm e}^{-\beta(-\frac{J}{2}-J_1)} \right]
\nonumber\\
\!&+\!\sinh\left(\frac{\beta h}{2}\right){\rm e}^{-\frac{\beta}{2}h}\left[-30{\rm e}^{-\beta(-2J+\frac{J_1}{2})}\!-\!20{\rm e}^{-\beta(-\frac{J}{2}+J_1)} \!-\!20{\rm e}^{-\beta(-\frac{J}{2}-J_1)} \right]
\nonumber\\
\!&+\!5{\rm e}^{-\beta(-\frac{J}{2}-2J_1)}\left[
3\cosh\left(\frac{3\beta}{2}(J\!-\!J_1)\right)\!+\!\sinh\left(\frac{3\beta}{2}(J\!-\!J_1)\right)
\right]
\bigg\},
\label{a7}\\
\rho_{10,10}\!&=\!\rho_{3,3}(-h),
\label{a8}\\
\rho_{4,4}\!&=\!\frac{{\rm e}^{ \frac{\beta }{4}J_1}}{180{\cal Z}}\bigg\{
\cosh\left(\frac{\beta h}{2}\right){\rm e}^{\frac{3}{2}\beta h}\left[25{\rm e}^{-\beta(J-\frac{J_1}{2})}\!+\!45{\rm e}^{-\beta(J+\frac{5}{2}J_1)}\!+\!180{\rm e}^{-\beta(-\frac{J}{2}+J_1)} \right]
\nonumber\\
\!&+\!\cosh\left(\frac{\beta h}{2}\right){\rm e}^{\frac{\beta}{2} h}\left[10{\rm e}^{-\beta(J-\frac{J_1}{2})} \!+\!80{\rm e}^{-\beta(-2J+\frac{J_1}{2})} \!+\!18{\rm e}^{-\beta(J+\frac{5}{2}J_1)} \!+\!17{\rm e}^{-\beta(J-\frac{5}{2}J_1)}\!+\!55{\rm e}^{-\beta(-\frac{J}{2}+J_1)}\right.
\nonumber\\
&\hspace*{3cm}\left.\!+\!65{\rm e}^{-\beta(-\frac{J}{2}-J_1)}\right]
\nonumber\\
\!&+\!\sinh\left(\frac{\beta h}{2}\right){\rm e}^{\frac{3}{2}\beta h}\left[35{\rm e}^{-\beta(J-\frac{J_1}{2})}\!+\!15{\rm e}^{-\beta(J+\frac{5}{2}J_1)}\!+\!60{\rm e}^{-\beta(-\frac{J}{2}+J_1)} \right]
\nonumber\\
\!&+\!\sinh\left(\frac{\beta h}{2}\right){\rm e}^{\frac{\beta}{2} h}\left[15{\rm e}^{-\beta(J-\frac{5}{2}J_1)}\!-\!25{\rm e}^{-\beta(-\frac{J}{2}+J_1)}\!-\!15{\rm e}^{-\beta(-\frac{J}{2}-J_1)} \right]
\nonumber\\
\!&+\!5{\rm e}^{-\beta(-\frac{J}{2}-2J_1)}
\left[9\cosh\left(\frac{3\beta}{2}(J\!-\!J_1)\right)\!+\!7\sinh\left(\frac{3\beta}{2}(J\!-\!J_1)\right)\right]\bigg\},
\label{a9}\\
\rho_{9,9}\!&=\!\rho_{4,4}(-h),
\label{a10}\\
\rho_{5,5}\!&=\!\frac{{\rm e}^{ \frac{\beta }{4}J_1}}{180{\cal Z}}\bigg\{
\cosh\left(\frac{\beta h}{2}\right){\rm e}^{\frac{\beta }{2}h}\left[80{\rm e}^{-\beta(J-\frac{J_1}{2})}\!+\!36{\rm e}^{-\beta(J+\frac{5}{2}J_1)}\!+\!24{\rm e}^{-\beta(J-\frac{5}{2}J_1)}\!+\!30{\rm e}^{-\beta(-\frac{J}{2}+J_1)}\right.
\nonumber\\
&\!\hspace*{4.5cm}+\!\left.110{\rm e}^{-\beta(-\frac{J}{2}-J_1)} \right]
\nonumber\\
\!&+\!\cosh\left(\frac{\beta h}{2}\right){\rm e}^{-\frac{\beta }{2}h}\left[50{\rm e}^{-\beta(-2J+\frac{J_1}{2})}\!+\!48{\rm e}^{-\beta(J+\frac{5}{2}J_1)}\!+\!32{\rm e}^{-\beta(J-\frac{5}{2}J_1)}\!+\!100{\rm e}^{-\beta(-\frac{J}{2}+J_1)} \right]
\nonumber\\
\!&+\!\sinh\left(\frac{\beta h}{2}\right){\rm e}^{\frac{\beta }{2}h}\left[40{\rm e}^{-\beta(J-\frac{J_1}{2})}\!+\!12{\rm e}^{-\beta(J+\frac{5}{2}J_1)}\!+\!48{\rm e}^{-\beta(J-\frac{5}{2}J_1)}\!-\!20{\rm e}^{-\beta(-\frac{J}{2}-J_1)} \right]
\nonumber\\
\!&+\!\sinh\left(\frac{\beta h}{2}\right){\rm e}^{-\frac{\beta }{2}h}\left[-30{\rm e}^{-\beta(-2J+\frac{J_1}{2})}\!-\!50{\rm e}^{-\beta(-\frac{J}{2}+J_1)}\!-\!50{\rm e}^{-\beta(-\frac{J}{2}-J_1)} \right]
\nonumber\\
\!&+\!10{\rm e}^{-\beta(-\frac{J}{2}-2J_1)}
\left[3\cosh\left(\frac{3\beta}{2}(J\!-\!J_1)\right)\!-\!\sinh\left(\frac{3\beta}{2}(J\!-\!J_1)\right)\right]\bigg\},
\label{a11}\\
\rho_{8,8}\!&=\!\rho_{5,5}(-h),
\label{a12}\\
\rho_{6,6}\!&=\!\frac{{\rm e}^{ \frac{\beta }{4}J_1}}{60{\cal Z}}\bigg\{
\cosh\left(\frac{\beta h}{2}\right){\rm e}^{-\frac{3}{2}\beta h}\left[20{\rm e}^{-\beta(J-\frac{J_1}{2})}\!+\!15{\rm e}^{-\beta(J+\frac{5}{2}J_1)}\!+\!45{\rm e}^{-\beta(-\frac{J}{2}+J_1)} \right]
\nonumber\\
\!&+\!\cosh\left(\frac{\beta h}{2}\right){\rm e}^{-\frac{\beta}{2}h}\left[25{\rm e}^{-\beta(J-\frac{J_1}{2})}\!+\!6{\rm e}^{-\beta(J+\frac{5}{2}J_1)}\!+\!39{\rm e}^{-\beta(J-\frac{5}{2}J_1)}\!+\!30{\rm e}^{-\beta(-\frac{J}{2}-J_1)} \right]
\nonumber\\
\!&+\!\sinh\left(\frac{\beta h}{2}\right){\rm e}^{-\frac{3}{2}\beta h}\left[-5{\rm e}^{-\beta(J+\frac{5}{2}J_1)}\!-\!35{\rm e}^{-\beta(-\frac{J}{2}+J_1)} \right]
\nonumber\\
\!&+\!\sinh\left(\frac{\beta h}{2}\right){\rm e}^{-\frac{\beta}{2}h}\left[5{\rm e}^{-\beta(J-\frac{J_1}{2})}\!+\!15{\rm e}^{-\beta(J-\frac{5}{2}J_1)} \right]
\nonumber\\
\!&-\!30{\rm e}^{-\frac{\beta}{4}(J-9J_1)}
\sinh\left(\frac{\beta}{4}(3J\!-\!5J_1)\right)\bigg\},
\label{a13}\\
\rho_{7,7}\!&=\!\rho_{6,6}(-h),
\label{a14}\\
\rho_{2,4}\!&=\!\rho_{4,2}\!=\!\frac{\sqrt{2}{\rm e}^{ \frac{\beta }{4}J_1}}{180{\cal Z}}\bigg\{
\cosh\left(\frac{\beta h}{2}\right){\rm e}^{\frac{3}{2}\beta h}\left[25{\rm e}^{-\beta(J-\frac{J_1}{2})}\!+\!45{\rm e}^{-\beta(J+\frac{5}{2}J_1)}\!-\! 90{\rm e}^{-\beta(-\frac{J}{2}+J_1)}\right]
\nonumber\\
\!&+\!\cosh\left(\beta h\right){\rm e}^{\frac{\beta }{2}h}\left[10{\rm e}^{-\beta(J-\frac{J_1}{2})}\!-\!40{\rm e}^{-\beta(-2J+\frac{J_1}{2})}\!+\!18{\rm e}^{-\beta(J+\frac{5}{2}J_1)}\!+\!17{\rm e}^{-\beta(J-\frac{5}{2}J_1)}\!+\! 25{\rm e}^{-\beta(-\frac{J}{2}+J_1)}\right.
\nonumber\\
&\hspace*{3cm}\left.\!+\! 5{\rm e}^{-\beta(-\frac{J}{2}-J_1)}\right]
\nonumber\\
\!&+\!\sinh\left(\frac{\beta h}{2}\right){\rm e}^{\frac{3}{2}\beta h}\left[35{\rm e}^{-\beta(J-\frac{J_1}{2})}\!+\!15{\rm e}^{-\beta(J+\frac{5}{2}J_1)}\!-\! 30{\rm e}^{-\beta(-\frac{J}{2}+J_1)}\right]
\nonumber\\
\!&+\!\sinh\left(\frac{\beta h}{2}\right){\rm e}^{\frac{\beta}{2}h}\left[15{\rm e}^{-\beta(J-\frac{5}{2}J_1)}\!+\! 5{\rm e}^{-\beta(-\frac{J}{2}+J_1)}\!-\! 15{\rm e}^{-\beta(-\frac{J}{2}-J_1)}\right]
\nonumber\\
\!&-\!5{\rm e}^{-\beta(-\frac{J}{2}-2J_1)}
\left[3\cosh\left(\frac{3\beta}{2}(J\!-\!J_1)\right)\!+\!5\sinh\left(\frac{3\beta}{2}(J\!-\!J_1)\right)\right]\bigg\},
\label{a15}\\
\rho_{11,9}\!&=\rho_{9,11}\!=\!\!\rho_{2,4}(-h),
\label{a16}\\
\rho_{2,7}\!&=\!\rho_{7,2}\!=\!-\frac{\sqrt{2}{\rm e}^{ \frac{\beta }{4}J_1}}{60{\cal Z}}\bigg\{
\cosh\left(\frac{\beta h}{2}\right){\rm e}^{\frac{3}{2}\beta h}\left[15{\rm e}^{-\beta(J-\frac{J_1}{2})}\!-\!15{\rm e}^{-\beta(J+\frac{5}{2}J_1)}\right]
\nonumber\\
\!&+\!\cosh\left(\beta h\right){\rm e}^{\frac{\beta}{2}h}\left[-10{\rm e}^{-\beta(J-\frac{J_1}{2})}\!-\!6{\rm e}^{-\beta(J+\frac{5}{2}J_1)}\!+\!11{\rm e}^{-\beta(J-\frac{5}{2}J_1)}\!+\! 10{\rm e}^{-\beta(-\frac{J}{2}+J_1)}\!-\!10{\rm e}^{-\beta(-\frac{J}{2}-J_1)}\right]
\nonumber\\
\!&+\!\sinh\left(\frac{\beta h}{2}\right){\rm e}^{\frac{3}{2}\beta h}\left[5{\rm e}^{-\beta(J-\frac{J_1}{2})}\!-\!5{\rm e}^{-\beta(J+\frac{5}{2}J_1)}\right]
\nonumber\\
\!&+\!\sinh\left(\frac{\beta h}{2}\right){\rm e}^{\frac{\beta}{2}h}\left[5{\rm e}^{-\beta(J-\frac{5}{2}J_1)}\right]
\nonumber\\
\!&-\!10{\rm e}^{-\frac{\beta}{4}(J-5J_1)}
\cosh\left(\frac{\beta}{4}(3J\!-\!9J_1)\right)\!+\!5{\rm e}^{-\frac{\beta}{4}(J-9J_1)}\left[3\cosh\left(\frac{\beta}{4}(3J\!-\!5J_1)\right)\!-\!\sinh\left(\frac{\beta}{4}(3J\!-\!5J_1)\right)\right]\bigg\},
\label{a17}\\
\rho_{11,6}\!&=\rho_{6,11}\!=\!\!\rho_{2,7}(-h),
\label{a18}\\
\rho_{4,7}\!&=\rho_{7,4}\!=\!\!\frac{\rho_{2,7}}{\sqrt{2}},
\label{a19}\\
\rho_{9,6}\!&=\rho_{6,9}\!=\!\!\frac{\rho_{2,7}(-h)}{\sqrt{2}},
\label{a20}\\
\rho_{3,5}\!&=\!\rho_{5,3}\!=\!\frac{\sqrt{2}{\rm e}^{ \frac{\beta }{4}J_1}}{180{\cal Z}}\bigg\{
\cosh\left(\frac{\beta h}{2}\right){\rm e}^{\frac{\beta }{2} h}\left[40{\rm e}^{-\beta(J-\frac{J_1}{2})}\!+\!18{\rm e}^{-\beta(J+\frac{5}{2}J_1)}\!+\!12{\rm e}^{-\beta(J-\frac{5}{2}J_1)}\!-\! 30{\rm e}^{-\beta(-\frac{J}{2}+J_1)}\right.
\nonumber\\
&\hspace*{5.5cm}\left.\!-\!50{\rm e}^{-\beta(-\frac{J}{2}-J_1)}\right]
\nonumber\\
\!&+\!\cosh\left(\frac{\beta h}{2}\right){\rm e}^{-\frac{\beta }{2} h}\left[-50{\rm e}^{-\beta(-2J+\frac{J_1}{2})}\!+\!24{\rm e}^{-\beta(J+\frac{5}{2}J_1)}\!+\!16{\rm e}^{-\beta(J-\frac{5}{2}J_1)}\!+\! 20{\rm e}^{-\beta(-\frac{J}{2}+J_1)}\right]
\nonumber\\
\!&+\!\sinh\left(\frac{\beta h}{2}\right){\rm e}^{\frac{\beta }{2} h}\left[20{\rm e}^{-\beta(J-\frac{J_1}{2})}\!+\!6{\rm e}^{-\beta(J+\frac{5}{2}J_1)}\!+\!24{\rm e}^{-\beta(J-\frac{5}{2}J_1)}\!-\! 40{\rm e}^{-\beta(-\frac{J}{2}-J_1)}\right]
\nonumber\\
\!&+\!\sinh\left(\frac{\beta h}{2}\right){\rm e}^{-\frac{\beta }{2} h}\left[30{\rm e}^{-\beta(-2J+\frac{J_1}{2})}\!-\! 10{\rm e}^{-\beta(-\frac{J}{2}+J_1)}\!-\!10{\rm e}^{-\beta(-\frac{J}{2}-J_1)}\right]
\nonumber\\
\!&-\!20{\rm e}^{-\beta(-\frac{J}{2}-2J_1)}
\sinh\left(\frac{3\beta}{2}(J\!-\!J_1)\right)\bigg\},
\label{a21}\\
\rho_{10,8}\!&=\rho_{8,10}\!=\!\!\rho_{3,5}(-h),
\label{a22}\\
\rho_{3,8}\!&=\!\rho_{8,3}\!=\!\rho_{10,5}\!=\!\rho_{5,10}\!=\!\frac{\sqrt{2}{\rm e}^{ \frac{\beta }{4}J_1}}{90{\cal Z}}\bigg\{
\cosh^2\left(\frac{\beta h}{2}\right)\left[21{\rm e}^{-\beta(J+\frac{5}{2}J_1)}\!-\! 11{\rm e}^{-\beta(J-\frac{5}{2}J_1)}\!+\! 20{\rm e}^{-\beta(-\frac{J}{2}+J_1)}\right.
\nonumber\\
&\!\hspace*{5.3cm}-\!\left.20{\rm e}^{-\beta(-\frac{J}{2}-J_1)}\right]
\nonumber\\
\!&+\!\sinh^2\left(\frac{\beta h}{2}\right)\left[3{\rm e}^{-\beta(J+\frac{5}{2}J_1)}\!-\!13{\rm e}^{-\beta(J-\frac{5}{2}J_1)}\!+\! 10{\rm e}^{-\beta(-\frac{J}{2}+J_1)}\!-\! 10{\rm e}^{-\beta(-\frac{J}{2}-J_1)}\right]
\nonumber\\
\!&+\!10{\rm e}^{-\beta(-\frac{J}{2}-2J_1)}
\sinh\left(\frac{3\beta}{2}(J\!-\!J_1)\right)\!-\!10{\rm e}^{\frac{\beta}{2}J}
\cosh\left(\frac{\beta}{2}(3J\!-\!J_1)\right)\bigg\},
\label{a23}\\
\rho_{3,10}\!&=\!\rho_{10,3}\!=\!\frac{{\rm e}^{ \frac{\beta }{4}J_1}}{90{\cal Z}}\bigg\{
\cosh^2\left(\frac{\beta h}{2}\right)\left[21{\rm e}^{-\beta(J+\frac{5}{2}J_1)}\!-\! 11{\rm e}^{-\beta(J-\frac{5}{2}J_1)}\!-\! 55{\rm e}^{-\beta(-\frac{J}{2}+J_1)}\!+\!55{\rm e}^{-\beta(-\frac{J}{2}-J_1)}\right]
\nonumber\\
\!&+\!\sinh^2\left(\frac{\beta h}{2}\right)\left[3{\rm e}^{-\beta(J+\frac{5}{2}J_1)}\!-\! 13{\rm e}^{-\beta(J-\frac{5}{2}J_1)}\!-\! 5{\rm e}^{-\beta(-\frac{J}{2}+J_1)}\!+\! 5{\rm e}^{-\beta(-\frac{J}{2}-J_1)}\right]
\nonumber\\
\!&-\!5{\rm e}^{-\beta(-\frac{J}{2}-2J_1)}
\left[3\cosh\left(\frac{3\beta}{2}(J\!-\!J_1)\right)\!+\!\sinh\left(\frac{3\beta}{2}(J\!-\!J_1)\right)\right]
\nonumber\\
\!&+\!5{\rm e}^{\frac{\beta}{2}J}\left[\cosh\left(\frac{\beta}{2}(3J\!-\!J_1)\right)\!+\! 3\sinh\left(\frac{\beta}{2}(3J\!-\!J_1)\right)\right]\bigg\},
\label{a24}\\
\rho_{5,8}\!&=\!\rho_{8,5}\!=\!\frac{{\rm e}^{ \frac{\beta }{4}J_1}}{90{\cal Z}}\bigg\{
\cosh^2\left(\frac{\beta h}{2}\right)\left[42{\rm e}^{-\beta(J+\frac{5}{2}J_1)}\!-\! 22{\rm e}^{-\beta(J-\frac{5}{2}J_1)}\!-\! 35{\rm e}^{-\beta(-\frac{J}{2}+J_1)}\!+\!35{\rm e}^{-\beta(-\frac{J}{2}-J_1)}\right]
\nonumber\\
\!&+\!\sinh^2\left(\frac{\beta h}{2}\right)\left[6{\rm e}^{-\beta(J+\frac{5}{2}J_1)}\!-\! 26{\rm e}^{-\beta(J-\frac{5}{2}J_1)}\!+\! 5{\rm e}^{-\beta(-\frac{J}{2}+J_1)}\!-\! 5{\rm e}^{-\beta(-\frac{J}{2}-J_1)}\right]
\nonumber\\
\!&-\!5{\rm e}^{-\beta(-\frac{J}{2}-2J_1)}
\left[3\cosh\left(\frac{3\beta}{2}(J\!-\!J_1)\right)\!-\!\sinh\left(\frac{3\beta}{2}(J\!-\!J_1)\right)\right]
\nonumber\\
\!&-\!5{\rm e}^{\frac{\beta}{2}J}\left[
\cosh\left(\frac{\beta}{2}(3J\!-\!J_1)\right)\!-\!3\sinh\left(\frac{\beta}{2}(3J\!-\!J_1)\right)\right]
\bigg\}.
\label{a25}
\end{flalign}
\end{widetext}
Subsequently, the partial transposition of the reduced density matrix  $\hat{\rho}_{\mu_{1}|\mu_{2}S_{2}}^{T_{\mu_{1}}}$,  transposed with respect the $\mu_{1}$ spin, has the following block-diagonal structure involving two $1\times 1$ matrices $\mathbf{Q}_1^{\mu_{1}}$, $\overline{\mathbf{Q}}_1^{\mu_{1}}\!=\!\mathbf{Q}_1^{\mu_{1}}(-h)$, two $3\times 3$ matrices $\mathbf{Q}_2^{\mu_1}$, $\overline{\mathbf{Q}}_2^{\mu_{1}}\!=\!\mathbf{Q}_2^{\mu_{1}}(-h)$ and one $4\times 4$ matrix $\mathbf{Q}_3^{\mu_{1}}$ with the following elements
\begin{align}
\allowdisplaybreaks
\mathbf{Q}_1^{\mu_{1}}\!&=\!\left(
\begin{array}{c}
 \rho_{6,6}
\end{array}\right),
\mathbf{Q}_2^{\mu_{1}}\!=\!\left(
\begin{array}{ccc}
 \rho_{1,1} & \rho_{2,7} & \rho_{4,7}\\
 \rho_{2,7} & \rho_{8,8} & \rho_{8,10}\\
 \rho_{4,7} & \rho_{8,10} & \rho_{10,10}\\
\end{array}\right),\;\;
\nonumber\\
\mathbf{Q}_3^{\mu_{1}}\!&=\!\left(
\begin{array}{cccc}
 \rho_{2,2} & \rho_{2,4} & \rho_{3,8}& \rho_{5,8}\\
 \rho_{2,4} & \rho_{4,4} & \rho_{3,10}& \rho_{3,8}\\
 \rho_{3,8} & \rho_{3,10} & \rho_{9,9}& \rho_{9,11}\\
  \rho_{5,8} & \rho_{3,8} & \rho_{9,11}& \rho_{11,11}
\end{array}\right).
\label{a26}
\end{align}
Due to the complexity of all diagonal blocks, respective eigenvalues are calculated numerically.  The sum of absolute values of all negative eigenvalues of $\mathbf{Q}_2^{\mu_{1}}$, $\overline{\mathbf{Q}}_2^{\mu_{1}}$ and  $\mathbf{Q}_3^{\mu_1}$ determines the respective bipartite negativity ${\cal N}_{\mu_1|\mu_2S_2}$ between a single spin $\mu_1$ and the spin dimer $\mu_2-S_2$.
\section{Bipartite entanglement ${\cal N}_{\mu_{2}|\mu_{1}S_{2}}$  between the single spin $\mu_2$ and the spin dimer $\mu_1-S_2$}
\label{App B}
The reduced density operator $\hat{\rho}_{\mu_{2}|\mu_{1}S_{2}}$ is identical with the reduced density operator $\hat{\rho}_{\mu_{1}|\mu_{2}S_{2}}$ (Eq.~\eqref{a1}) and of course, the corresponding reduced density matrix is the same as that presented in Eq.~\eqref{a2}. A partial transpose of the density matrix~\eqref{a2} over a different spin, namely the spin $\mu_{2}$, leads again to the  block-diagonal structure with   two $1\times 1$ matrices $\mathbf{Q}_1^{\mu_{2}}$, $\overline{\mathbf{Q}}_1^{\mu_{2}}\!=\!\mathbf{Q}_1^{\mu_{2}}(-h)$, two $3\times 3$ matrices $\mathbf{Q}_2^{\mu_{2}}$, $\overline{\mathbf{Q}}_2^{\mu_{2}}\!=\!\mathbf{Q}_2^{\mu_{2}}(-h)$ and one $4\times 4$ matrix $\mathbf{Q}_3^{\mu_{2}}$. Nevertheless, the structure of each block matrix  is  different. In particular,
\begin{align}
\allowdisplaybreaks
\mathbf{Q}_1^{\mu_{2}}\!&=\!\left(
\begin{array}{c}
 \rho_{4,4}
\end{array}\right),
\mathbf{Q}_2^{\mu_{2}}\!=\!\left(
\begin{array}{ccc}
 \rho_{1,1} & \rho_{2,4} & \rho_{4,7}\\
 \rho_{2,4} & \rho_{5,5} & \rho_{3,8}\\
 \rho_{4,7} & \rho_{3,8} & \rho_{10,10}\\
\end{array}\right),\;\;
\nonumber\\
\mathbf{Q}_3^{\mu_{2}}\!&=\!\left(
\begin{array}{cccc}
 \rho_{2,2} & \rho_{2,7} & \rho_{3,5}& \rho_{5,8}\\
 \rho_{2,7} & \rho_{7,7} & \rho_{3,10}& \rho_{8,10}\\
 \rho_{3,5} & \rho_{3,10} & \rho_{6,6}& \rho_{6,11}\\
  \rho_{5,8} & \rho_{8,10} & \rho_{6,11}& \rho_{11,11}
\end{array}\right).
\label{b1}
\end{align}
Particular elements $\rho_{i,j}$ are the same as that presented in Eqs.~\eqref{a3}-\eqref{a25}. The sum of absolute values of all negative eigenvalues of $\mathbf{Q}_2^{\mu_{2}}$, $\overline{\mathbf{Q}}_2^{\mu_{2}}$ and  $\mathbf{Q}_3^{\mu_2}$ determines the respective bipartite negativity ${\cal N}_{\mu_2|\mu_1S_2}$ between a single spin $\mu_2$ and the spin dimer $\mu_1-S_2$.

\section{Bipartite entanglement ${\cal N}_{S_{1}|\mu_{1}\mu_{2}}$ between the single spin $S_1$ and the spin dimer $\mu_1-\mu_2$}
\label{App C}

The reduced density operator $\hat{\rho}_{S_{1}|\mu_{1}\mu_{2}}$ is identical with the reduced density operator $\hat{\rho}_{\mu_{1}|\mu_{2}S_{2}}$ (Eq.~\eqref{a1}) and of course, the corresponding reduced density matrix is the same as that presented in Eq.~\eqref{a2}. A partial transpose of the density matrix~\eqref{a2} over a different spin, namely the spin $S_{1}$, leads again to the  block-diagonal structure with   two $1\times 1$ matrices $\mathbf{Q}_1^{S_{1}}$, $\overline{\mathbf{Q}}_1^{S_{1}}\!=\!\mathbf{Q}_1^{S_{1}}(-h)$, two $3\times 3$ matrices $\mathbf{Q}_2^{S_{1}}$, $\overline{\mathbf{Q}}_2^{S_{1}}\!=\!\mathbf{Q}_2^{S_{1}}(-h)$ and one $4\times 4$ matrix $\mathbf{Q}_3^{S_{1}}$. Nevertheless, the structure of each block matrix  is  different. In particular,
\begin{align}
\allowdisplaybreaks
\mathbf{Q}_1^{S_{1}}\!&=\!\left(
\begin{array}{c}
 \rho_{3,3}
\end{array}\right),\;\;
\mathbf{Q}_2^{S_{1}}\!=\!\left(
\begin{array}{ccc}
 \rho_{2,2} & \rho_{3,5} & \rho_{3,8}\\
 \rho_{3,5} & \rho_{6,6} & \rho_{6,9}\\
 \rho_{3,8} & \rho_{6,9} & \rho_{9,9}\\
\end{array}\right),\;\;
\nonumber\\
\mathbf{Q}_3^{S_{1}}\!&=\!\left(
\begin{array}{cccc}
 \rho_{1,1} & \rho_{2,4} & \rho_{2,7}& \rho_{3,10}\\
 \rho_{2,4} & \rho_{5,5} & \rho_{5,8}& \rho_{6,11}\\
 \rho_{2,7} & \rho_{5,8} & \rho_{8,8}& \rho_{9,11}\\
  \rho_{3,10} & \rho_{6,11} & \rho_{9,11}& \rho_{12,12}
\end{array}\right).
\label{h1}
\end{align}
Particular elements $\rho_{i,j}$ are the same as that presented in Eqs.~\eqref{a3}-\eqref{a25}. The sum of absolute values of all negative eigenvalues of $\mathbf{Q}_2^{S_{1}}$, $\overline{\mathbf{Q}}_2^{S_{1}}$ and  $\mathbf{Q}_3^{S_1}$ determines the respective bipartite negativity ${\cal N}_{S_2|\mu_1\mu_2}$ between a single spin $S_2$ and the spin dimer $\mu_1-\mu_2$.
\\
\section{Bipartite entanglement ${\cal N}_{\mu_{1}|S_{1}S_{2}}$  between the single spin $\mu_1$ and the spin dimer $S_1-S_2$}
\label{App D}
The reduced density operator $\hat{\rho}_{\mu_{1}|S_{1}S_{2}}$ is defined after tracing out degree of freedom of the spin $\mu_{2}$. Thus,
\begin{align}
\hat{\rho}_{\mu_{1}|S_{1}S_{2}}&\!=\!\sum_{\mu_{2}^z} \langle \mu_{2}^z\vert\hat{\rho}\vert \mu_{2}^z\rangle
\nonumber\\
&\!=\!\frac{1}{\cal Z}\sum_{k=1}^{36}  {\rm e}^{-\beta \varepsilon_k}\left(\sum_{\mu_2^z} \langle \mu_2^z|\psi_k\rangle\langle\psi_k|\mu_2^z\rangle \right).
\label{d1}
\end{align}
 The corresponding reduced density matrix $\hat{\rho}_{\mu_{1}|S_{1}S_{2}}$ in the basis of $\vert \mu^z_{1},S_{1}^z,S^z_{2}\rangle$ reads as follows
\\\\\\
\begin{widetext}
\begin{align}
\allowdisplaybreaks
{\hat{\rho}_{\mu_{1}|S_{1},S_{2}}}\!=\!
\resizebox{0.9\textwidth}{!}{$
\begin{blockarray}{r ccc ccc ccc ccc ccc ccc}
 & \vert \frac{1}{2},1,1\rangle & \vert \frac{1}{2},1,0\rangle &\vert \frac{1}{2},1,\!-\!1\rangle  
 & \vert \frac{1}{2},0,1\rangle & \vert \frac{1}{2},0,0\rangle &\vert \frac{1}{2},0,\!-\!1\rangle 
 & \vert \frac{1}{2},\!-\!1,1\rangle & \vert \frac{1}{2},\!-\!1,0\rangle &\vert \frac{1}{2},\!-\!1,\!-\!1\rangle  
& \vert \frac{1}{2},1,1\rangle & \vert \!-\!\frac{1}{2},1,0\rangle &\vert \!-\!\frac{1}{2},1,\!-\!1\rangle  
 & \vert \frac{1}{2},0,1\rangle & \vert \!-\!\frac{1}{2},0,0\rangle &\vert \!-\!\frac{1}{2},0,\!-\!1\rangle 
 & \vert \frac{1}{2},\!-\!1,1\rangle & \vert \!-\!\frac{1}{2},\!-\!1,0\rangle &\vert \!-\!\frac{1}{2},\!-\!1,\!-\!1\rangle  
  \\
\begin{block}{r(ccc ccc ccc ccc ccc ccc)}
\langle \frac{1}{2}, 1,1\vert \;\;\;&\rho_{1,1} & 0 & 0 & 0 & 0 & 0& 0 & 0 & 0& 0 & 0 & 0 & 0 & 0 & 0 & 0 & 0 & 0 \\
\langle \frac{1}{2}, 1,0\vert \;\;\;&0 & \rho_{2,2} & 0 & \rho_{2,4} & 0 & 0 & 0 & 0 & 0 & \rho_{2,10} & 0 & 0 & 0 & 0 & 0& 0 & 0 & 0 \\
\langle \frac{1}{2}, 1,\!-\!1\vert\;\;\; & 0 & 0 & \rho_{3,3} &  0 &\rho_{3,5}& 0&  \rho_{3,7}& 0 & 0 & 0 & \rho_{3,11} & 0 &\rho_{3,13}& 0 & 0 & 0& 0 & 0 \\
 \langle \!-\!\frac{1}{2},0,1\vert\; \;\;&0 & \rho_{4,2} & 0 & \rho_{4,4} & 0 & 0 & 0 & 0 & 0 & \rho_{4,10} & 0 & 0 & 0 & 0 & 0& 0 & 0 & 0 \\
\langle \!-\!\frac{1}{2},0,0\vert\;\;\;& 0 & 0 & \rho_{5,3} &  0 &\rho_{5,5}& 0&  \rho_{5,7}& 0 & 0 & 0 & \rho_{5,11} & 0 &\rho_{5,13}& 0 & 0 & 0& 0 & 0 \\
 \langle \!-\!\frac{1}{2},0,\!-\!1\vert \;\;\;& 0 & 0 & 0 & 0 & 0 & \rho_{6,6}& 0 & \rho_{6,8} & 0 & 0 & 0 & \rho_{6,12}& 0& \rho_{6,14}& 0 & \rho_{6,16}  & 0 & 0 \\
\langle \frac{1}{2}, \!-\!1,1\vert \;\;\;& 0 & 0 & \rho_{7,3} &  0 &\rho_{7,5}& 0&  \rho_{7,7}& 0 & 0 & 0 & \rho_{7,11} & 0 &\rho_{7,13}& 0 & 0 & 0& 0 & 0 \\
\langle \frac{1}{2}, \!-\!1,0\vert \;\;\;& 0 & 0 & 0 & 0 & 0 & \rho_{8,6}& 0 & \rho_{8,8} & 0 & 0 & 0 & \rho_{8,12}& 0& \rho_{8,14}& 0 & \rho_{8,16}  & 0 & 0 \\
\langle \frac{1}{2}, \!-\!1,\!-\!1\vert\;\;\; & 0 & 0& 0 & 0& 0 & 0& 0 & 0 & \rho_{9,9} &  0 & 0 & 0& 0 & 0&\rho_{9,15}& 0&  \rho_{9,17}& 0 \\
 \langle \!-\!\frac{1}{2},1,1\vert \;\;\;&0 & \rho_{10,2} & 0 & \rho_{10,4} & 0 & 0 & 0 & 0 & 0 & \rho_{10,10} & 0 & 0 & 0 & 0 & 0& 0 & 0 & 0 \\
  \langle \!-\!\frac{1}{2}, 1,0\vert \;\;\;& 0 & 0 & \rho_{11,3} &  0 &\rho_{11,5}& 0&  \rho_{11,7}& 0 & 0 & 0 & \rho_{11,11} & 0 &\rho_{11,13}& 0 & 0 & 0& 0 & 0 \\
  \langle \!-\!\frac{1}{2}, 1,\!-\!1\vert \;\;\;& 0 & 0 & 0 & 0 & 0 & \rho_{12,6}& 0 & \rho_{12,8} & 0 & 0 & 0 & \rho_{12,12}& 0& \rho_{12,14}& 0 & \rho_{12,16}  & 0 & 0 \\
  \langle \!-\!\frac{1}{2}, 0,1\vert \;\;\;& 0 & 0 & \rho_{13,3} &  0 &\rho_{13,5}& 0&  \rho_{13,7}& 0 & 0 & 0 & \rho_{13,11} & 0 &\rho_{13,13}& 0 & 0 & 0& 0 & 0 \\
  \langle \!-\!\frac{1}{2}, 0,0\vert \;\;\;& 0 & 0 & 0 & 0 & 0 & \rho_{14,6}& 0 & \rho_{14,8} & 0 & 0 & 0 & \rho_{14,12}& 0& \rho_{14,14}& 0 & \rho_{14,16}  & 0 & 0 \\
  \langle \!-\!\frac{1}{2}, 0,\!-\!1\vert \;\;\;& 0 & 0& 0 & 0& 0 & 0& 0 & 0 & \rho_{15,9} &  0 & 0 & 0& 0 & 0&\rho_{15,15}& 0&  \rho_{15,17}& 0 \\
   \langle \!-\!\frac{1}{2}, \!-\!1,1\vert \;\;\;& 0 & 0 & 0 & 0 & 0 & \rho_{16,6}& 0 & \rho_{16,8} & 0 & 0 & 0 & \rho_{16,12}& 0& \rho_{16,14}& 0 & \rho_{16,16}  & 0 & 0 \\
   \langle \!-\!\frac{1}{2}, \!-\!1,0\vert \;\;\;& 0 & 0& 0 & 0& 0 & 0& 0 & 0 & \rho_{17,9} &  0 & 0 & 0& 0 & 0&\rho_{17,15}& 0&  \rho_{17,17}& 0 \\
    \langle \!-\!\frac{1}{2}, \!-\!1,\!-\!1\vert \;\;\;& 0 & 0& 0 & 0& 0 & 0& 0 & 0& 0 & 0& 0 & 0& 0 & 0& 0 & 0& 0 & \rho_{18,18}
 \\
\end{block}
\end{blockarray}\;\;,$}
\label{d2}
\end{align}
where
\begin{flalign}
\rho_{1,1}\!&=\!\frac{{\rm e}^{ \frac{\beta }{4}J_1}}{6{\cal Z}}\bigg\{
\cosh\left(\frac{\beta h}{2}\right){\rm e}^{\frac{5}{2}\beta h}\left[7{\rm e}^{-\beta(J+\frac{5}{2}J_1)} \right]
\nonumber\\
\!&+\!\sinh\left(\frac{\beta h}{2}\right){\rm e}^{\frac{5}{2}\beta h}\left[5{\rm e}^{-\beta(J+\frac{5}{2}J_1)}\right]
\nonumber\\
\!&+\!{\rm e}^{-\frac{\beta}{4}(J+J_1)}{\rm e}^{2\beta h}\left[ 
5\cosh\left(\frac{3\beta}{4}(J\!-\!J_1)\right)\right.
\!+\!\left.3\sinh\left(\frac{3\beta}{4}(J\!-\!J_1)\right)\right]\bigg\},
\label{d3}\\
\rho_{18,18}\!&=\!\rho_{1,1}(-h),
\label{d4}\\
\rho_{2,2}\!&=\!\frac{{\rm e}^{ \frac{\beta }{4}J_1}}{60{\cal Z}}\bigg\{
\cosh\left(\frac{\beta h}{2}\right){\rm e}^{\frac{3}{2}\beta h}\left[40{\rm e}^{-\beta(J-\frac{J_1}{2})}\!+\!28{\rm e}^{-\beta(J+\frac{5}{2}J_1)}\!+\!25{\rm e}^{-\beta(-\frac{J}{2}+J_1)} \right]
\nonumber\\
\!&+\!\sinh\left(\frac{\beta h}{2}\right){\rm e}^{\frac{3}{2}\beta h}\left[12{\rm e}^{-\beta(J+\frac{5}{2}J_1)}\!+\!15{\rm e}^{-\beta(-\frac{J}{2}+J_1)} \right]
\nonumber\\
\!&+\!3{\rm e}^{-\frac{\beta}{4}(J-7J_1)}{\rm e}^{\beta h}\left[
9\cosh\left(\frac{3\beta}{4}(J\!-\!J_1)\right)\!+\!\sinh\left(\frac{3\beta}{4}(J\!-\!J_1)\right)
\right]
\bigg\},
\label{d5}\\
\rho_{17,17}\!&=\!\rho_{2,2}(-h),
\label{d6}\\
\rho_{3,3}\!&=\!\frac{{\rm e}^{ \frac{\beta }{4}J_1}}{60{\cal Z}}\bigg\{
\cosh\left(\frac{\beta h}{2}\right){\rm e}^{\frac{\beta }{2}h}\left[25{\rm e}^{-\beta(J-\frac{J_1}{2})}\!+\!7{\rm e}^{-\beta(J+\frac{5}{2}J_1)}\!+\!33{\rm e}^{-\beta(J-\frac{5}{2}J_1)}\!+\!20{\rm e}^{-\beta(-\frac{J}{2}+J_1)}\right.
\nonumber\\
&\!\hspace*{4.5cm}+\!\left.60{\rm e}^{-\beta(-\frac{J}{2}-J_1)} \right]
\nonumber\\
\!&+\!\sinh\left(\frac{\beta h}{2}\right){\rm e}^{\frac{\beta }{2}h}\left[-5{\rm e}^{-\beta(J-\frac{J_1}{2})}\!+\!{\rm e}^{-\beta(J+\frac{5}{2}J_1)}\!-\!21{\rm e}^{-\beta(J-\frac{5}{2}J_1)}\right]
\nonumber\\
\!&-\!5{\rm e}^{-\frac{\beta}{4}(J-9J_1)}\left[
\cosh\left(\frac{\beta}{4}(3J\!-\!5J_1)\right)\!+\!11\sinh\left(\frac{\beta}{4}(3J\!-\!5J_1)\right)
\right]
\nonumber\\
\!&-\!20{\rm e}^{-\frac{\beta}{4}(J-5J_1)}\cosh\left(\frac{3\beta}{4}(J\!-\!3J_1)\right)
\bigg\},
\label{d7}\\
\rho_{16,16}\!&=\!\rho_{3,3}(-h),
\label{d8}\\
\rho_{4,4}\!&=\!\frac{{\rm e}^{ \frac{\beta }{4}J_1}}{180{\cal Z}}\bigg\{
\cosh\left(\frac{\beta h}{2}\right){\rm e}^{\frac{3}{2}\beta h}\left[120{\rm e}^{-\beta(J-\frac{J_1}{2})}\!+\!84{\rm e}^{-\beta(J+\frac{5}{2}J_1)}\!+\!135{\rm e}^{-\beta(-\frac{J}{2}+J_1)} \right]
\nonumber\\
\!&+\!\sinh\left(\frac{\beta h}{2}\right){\rm e}^{\frac{3}{2}\beta h}\left[36{\rm e}^{-\beta(J+\frac{5}{2}J_1)}\!-\!15{\rm e}^{-\beta(-\frac{J}{2}+J_1)} \right]
\nonumber\\
\!&+\!5{\rm e}^{-\frac{\beta}{4}(-5J-J_1)}{\rm e}^{\beta h}
\left[13\cosh\left(\frac{3\beta}{4}(J\!-\!J_1)\right)\!+\!3\sinh\left(\frac{3\beta}{4}(J\!-\!J_1)\right)\right]
\nonumber\\
\!&-\!4{\rm e}^{-\beta(J-\frac{3}{2}J_1)}{\rm e}^{\beta h}
\left[11\cosh\left(\beta J_1\right)\!-\!19\sinh\left(\beta J_1\right)\right]
\bigg\},
\label{d9}\\
\rho_{15,15}\!&=\!\rho_{4,4}(-h),
\label{d10}\\
\rho_{5,5}\!&=\!\frac{{\rm e}^{ \frac{\beta }{4}J_1}}{90{\cal Z}}\bigg\{
\cosh\left(\frac{\beta h}{2}\right){\rm e}^{\frac{\beta }{2}h}\left[15{\rm e}^{-\beta(-2J+\frac{J_1}{2})}\!+\!42{\rm e}^{-\beta(J+\frac{5}{2}J_1)}\!+\!18{\rm e}^{-\beta(J-\frac{5}{2}J_1)}\!+\!50{\rm e}^{-\beta(-\frac{J}{2}+J_1)}\right.
\nonumber\\
&\!\hspace*{4.5cm}+\!\left.30{\rm e}^{-\beta(-\frac{J}{2}-J_1)} \right]
\nonumber\\
\!&+\!\sinh\left(\frac{\beta h}{2}\right){\rm e}^{\frac{\beta }{2}h}\left[5{\rm e}^{-\beta(-2J+\frac{J_1}{2})}\!+\!6{\rm e}^{-\beta(J+\frac{5}{2}J_1)}\!+\!14{\rm e}^{-\beta(J-\frac{5}{2}J_1)}\!+\!10{\rm e}^{-\beta(-\frac{J}{2}+J_1)}\!-\!10{\rm e}^{-\beta(-\frac{J}{2}-J_1)} \right]
\nonumber\\
\!&+\!10{\rm e}^{-\beta(-\frac{J}{2}-2J_1)}
\cosh\left(\frac{3\beta}{2}(J\!-\!J_1)\right)
\!+\!5{\rm e}^{-\beta(J-2J_1)}{\rm e}^{\beta h}
\left[3\cosh\left(\frac{3}{2}\beta J_1\right)\!-\!\sinh\left(\frac{3}{2}\beta J_1\right)\right]
\bigg\},
\label{d11}\\
\rho_{14,14}\!&=\!\rho_{5,5}(-h),
\label{d12}\\
\rho_{6,6}\!&=\!\frac{{\rm e}^{ \frac{\beta }{4}J_1}}{180{\cal Z}}\bigg\{
\cosh\left(\frac{\beta h}{2}\right){\rm e}^{-\frac{\beta}{2}h}\left[70{\rm e}^{-\beta(J-\frac{J_1}{2})}\!+\!42{\rm e}^{-\beta(J+\frac{5}{2}J_1)}\!+\!38{\rm e}^{-\beta(J-\frac{5}{2}J_1)}\!+\!65{\rm e}^{-\beta(-\frac{J}{2}+J_1)} \right.
\nonumber\\
&\!\hspace*{4.9cm}+\!\left.95{\rm e}^{-\beta(-\frac{J}{2}-J_1)} \right]
\nonumber\\
\!&+\!\sinh\left(\frac{\beta h}{2}\right){\rm e}^{-\frac{\beta}{2}h}\left[-50{\rm e}^{-\beta(J-\frac{J_1}{2})}\!-\!6{\rm e}^{-\beta(J+\frac{5}{2}J_1)}\!-\!34{\rm e}^{-\beta(J-\frac{5}{2}J_1)} \!+\!35{\rm e}^{-\beta(-\frac{J}{2}+J_1)} \!+\!5{\rm e}^{-\beta(-\frac{J}{2}-J_1)} \right]
\nonumber\\
\!&+\!20{\rm e}^{-\frac{\beta}{2}(-J-3J_1)}
\cosh\left(\frac{\beta}{2}(3J\!-\!4J_1)\right)\!+\!10{\rm e}^{2\beta J}\left[3\cosh\left(\frac{\beta}{2} J_1\right)\!+\! \sinh\left(\frac{\beta}{2} J_1\right)\right]
\bigg\},
\label{d13}\\
\rho_{13,13}\!&=\!\rho_{6,6}(-h),
\label{d14}\\
\rho_{7,7}\!&=\!\frac{{\rm e}^{ \frac{\beta }{4}J_1}}{180{\cal Z}}\bigg\{
\cosh\left(\frac{\beta h}{2}\right){\rm e}^{\frac{\beta}{2}h}\left[35{\rm e}^{-\beta(J-\frac{J_1}{2})}\!+\!21{\rm e}^{-\beta(J+\frac{5}{2}J_1)}\!+\!19{\rm e}^{-\beta(J-\frac{5}{2}J_1)}\!+\!70{\rm e}^{-\beta(-\frac{J}{2}+J_1)} \right.
\nonumber\\
&\!\hspace*{4.5cm}+\!\left.130{\rm e}^{-\beta(-\frac{J}{2}-J_1)} \right]
\nonumber\\
\!&+\!\sinh\left(\frac{\beta h}{2}\right){\rm e}^{\frac{\beta}{2}h}\left[25{\rm e}^{-\beta(J-\frac{J_1}{2})}\!+\!3{\rm e}^{-\beta(J+\frac{5}{2}J_1)}\!+\!17{\rm e}^{-\beta(J-\frac{5}{2}J_1)} \!-\!10{\rm e}^{-\beta(-\frac{J}{2}+J_1)} \!+\!50{\rm e}^{-\beta(-\frac{J}{2}-J_1)} \right]
\nonumber\\
\!&+\!5{\rm e}^{-\frac{\beta}{2}(-J-3J_1)}
\left[5\cosh\left(\frac{\beta}{2}(3J\!-\!4J_1)\right)\!+\!3\sinh\left(\frac{\beta}{2}(3J\!-\!4J_1)\right)\right]
\nonumber\\
\!&+\!20{\rm e}^{2\beta J}\left[3\cosh\left(\frac{\beta}{2} J_1\right)\!+\! \sinh\left(\frac{\beta}{2} J_1\right)\right]
\bigg\},
\label{d15}\\
\rho_{12,12}\!&=\!\rho_{7,7}(-h),
\label{d16}\\
\rho_{8,8}\!&=\!\frac{{\rm e}^{ \frac{\beta }{4}J_1}}{180{\cal Z}}\bigg\{
\cosh\left(\frac{\beta h}{2}\right){\rm e}^{-\frac{\beta}{2}h}\left[60{\rm e}^{-\beta(-2J+\frac{J_1}{2})}\!+\!42{\rm e}^{-\beta(J+\frac{5}{2}J_1)}\!+\!18{\rm e}^{-\beta(J-\frac{5}{2}J_1)}\!+\!125{\rm e}^{-\beta(-\frac{J}{2}+J_1)} \right.
\nonumber\\
&\!\hspace*{4.9cm}+\!\left.75{\rm e}^{-\beta(-\frac{J}{2}-J_1)} \right]
\nonumber\\
\!&+\!\sinh\left(\frac{\beta h}{2}\right){\rm e}^{-\frac{\beta}{2}h}\left[-20{\rm e}^{-\beta(-2J+\frac{J_1}{2})}\!-\!6{\rm e}^{-\beta(J+\frac{5}{2}J_1)}\!-\!14{\rm e}^{-\beta(J-\frac{5}{2}J_1)} \!-\!25{\rm e}^{-\beta(-\frac{J}{2}+J_1)} \!+\!25{\rm e}^{-\beta(-\frac{J}{2}-J_1)} \right]
\nonumber\\
\!&+\!5{\rm e}^{-\beta(-\frac{J}{2}-2J_1)}
\left[5\cosh\left(\frac{3\beta}{2}(J\!-\!J_1)\right)\!+\!3\sinh\left(\frac{3\beta}{2}(J\!-\!J_1)\right)\right]
\nonumber\\
\!&+\!5{\rm e}^{-\beta (J-2J_1)}\left[3\cosh\left(\frac{3}{2}\beta J_1\right)\!-\! \sinh\left(\frac{3}{2}\beta J_1\right)\right]
\bigg\},
\label{d17}\\
\rho_{11,11}\!&=\!\rho_{8,8}(-h),
\label{d18}\\
\rho_{9,9}\!&=\!\frac{{\rm e}^{ \frac{\beta }{4}J_1}}{90{\cal Z}}\bigg\{
\cosh\left(\frac{\beta h}{2}\right){\rm e}^{-\frac{3}{2}\beta h}\left[30{\rm e}^{-\beta(J-\frac{J_1}{2})}\!+\!21{\rm e}^{-\beta(J+\frac{5}{2}J_1)}\!+\!90{\rm e}^{-\beta(-\frac{J}{2}+J_1)}\right]
\nonumber\\
\!&+\!\sinh\left(\frac{\beta h}{2}\right){\rm e}^{-\frac{3}{2}\beta h}\left[-9{\rm e}^{-\beta(J+\frac{5}{2}J_1)}\!-\!30{\rm e}^{-\beta(-\frac{J}{2}+J_1)} \right]
\nonumber\\
\!&+\!10{\rm e}^{-\frac{\beta}{4}(-5J-J_1)}{\rm e}^{-\beta h}
\left[5\cosh\left(\frac{3\beta}{4}(J\!-\!J_1)\right)\!+\!3\sinh\left(\frac{3\beta}{4}(J\!-\!J_1)\right)\right]
\nonumber\\
\!&-\!{\rm e}^{-\beta (J-\frac{3}{2}J_1)}{\rm e}^{-\beta h}
\left[11\cosh\left(\beta J_1\right)\!-\! 19\sinh\left(\beta J_1\right)\right]
\bigg\},
\label{d19}\\
\rho_{10,10}\!&=\!\rho_{9,9}(-h),
\label{d20}\\
\rho_{2,4}\!&=\!\rho_{4,2}\!=\!-\frac{{\rm e}^{ \frac{\beta }{4}J_1}}{30{\cal Z}}\bigg\{
\cosh\left(\frac{\beta h}{2}\right){\rm e}^{\frac{3}{2}\beta h}\left[20{\rm e}^{-\beta(J-\frac{J_1}{2})}\!-\!14{\rm e}^{-\beta(J+\frac{5}{2}J_1)}\right]
\nonumber\\
\!&+\!\sinh\left(\frac{\beta h}{2}\right){\rm e}^{\frac{3}{2}\beta h}\left[-6{\rm e}^{-\beta(J+\frac{5}{2}J_1)}\right]
\nonumber\\
\!&+\!10{\rm e}^{\frac{\beta}{2} J}{\rm e}^{\beta h}\sinh\left(\beta J_1\right)
\!-\!2{\rm e}^{-\beta (J-\frac{3}{2}J_1)}{\rm e}^{\beta h}
\left[3\cosh\left(\beta J_1\right)\!-\! 7\sinh\left(\beta J_1\right)\right]
\bigg\},
\label{d21}\\
\rho_{15,17}\!&=\rho_{17,15}\!=\!\!\rho_{2,4}(-h),
\label{d22}\\
\rho_{2,10}\!&=\rho_{10,2}\!=\!\!\frac{\sqrt{2}}{2}\rho_{2,4},
\label{d23}\\
\rho_{9,17}\!&=\rho_{17,9}\!=\!\!\frac{\sqrt{2}}{2}\rho_{2,4}(-h),
\label{d24}\\
\rho_{4,10}\!&=\!\rho_{10,4}\!=\!\frac{\sqrt{2}{\rm e}^{ \frac{\beta }{4}J_1}}{180{\cal Z}}\bigg\{
\cosh\left(\frac{\beta h}{2}\right){\rm e}^{\frac{3}{2}\beta h}\left[60{\rm e}^{-\beta(J-\frac{J_1}{2})}\!+\!42{\rm e}^{-\beta(J+\frac{5}{2}J_1)}\!-\! 45{\rm e}^{-\beta(-\frac{J}{2}+J_1)}\right]
\nonumber\\
\!&+\!\sinh\left(\frac{\beta h}{2}\right){\rm e}^{\frac{3}{2}\beta h}\left[18{\rm e}^{-\beta(J+\frac{5}{2}J_1)}\!-\! 75{\rm e}^{-\beta(-\frac{J}{2}+J_1)}\right]
\nonumber\\
\!&-\!5{\rm e}^{-\frac{\beta}{4} (-5J-J_1)}{\rm e}^{\beta h}\left[ 7\cosh\left(\frac{3\beta}{4}(J\!-\!J_1)\right)\!+\!9\sinh\left(\frac{3\beta}{4}(J\!-\!J_1)\right)\right]
\nonumber\\
\!&-\!2{\rm e}^{-\beta (J-\frac{3}{2}J_1)}{\rm e}^{\beta h}
\left[11\cosh\left(\beta J_1\right)\!-\! 19\sinh\left(\beta J_1\right)\right]
\bigg\},
\label{d25}\\
\rho_{9,15}\!&=\rho_{15,9}\!=\!\!\rho_{4,10}(-h),
\label{d26}\\
\rho_{3,7}\!&=\!\rho_{7,3}\!=\!-\frac{{\rm e}^{ \frac{\beta }{24}J_1}}{60{\cal Z}}\bigg\{
\cosh\left(\frac{\beta h}{2}\right){\rm e}^{\frac{\beta}{2}h }\left[15{\rm e}^{-\beta(J-\frac{J_1}{2})}\!-\!8{\rm e}^{-\beta(J+\frac{5}{2}J_1)}\!-\!3{\rm e}^{-\beta(J-\frac{5}{2}J_1)}\right]
\nonumber\\
\!&+\!\sinh\left(\frac{\beta h}{2}\right){\rm e}^{\frac{\beta}{2}h}\left[5{\rm e}^{-\beta(J-\frac{J_1}{2})}\!-\!9{\rm e}^{-\beta(J-\frac{5}{2}J_1)}\right]
\nonumber\\
\!&-\!{\rm e}^{-\beta (J-\frac{J_1}{2})}
\left[4\cosh\left(3\beta J_1\right)\!+\! 6\sinh\left(3\beta J_1\right)\right]
\bigg\},
\label{d27}\\
\rho_{12,16}\!&=\rho_{16,12}\!=\!\!\rho_{3,7}(-h),
\label{d28}\\
\rho_{3,13}\!&=\rho_{13,3}\!=\!\!\sqrt{2}\rho_{3,7},
\label{d29}\\
\rho_{6,16}\!&=\rho_{16,6}\!=\!\!\sqrt{2}\rho_{3,7}(-h),
\label{d30}\\
\rho_{7,13}\!&=\!\rho_{13,7}\!=\!\frac{\sqrt{2}{\rm e}^{ \frac{\beta }{4}J_1}}{180{\cal Z}}\bigg\{
\cosh\left(\frac{\beta h}{2}\right){\rm e}^{\frac{\beta}{2}h }\left[35{\rm e}^{-\beta(J-\frac{J_1}{2})}\!+\!21{\rm e}^{-\beta(J+\frac{5}{2}J_1)}\!+\!19{\rm e}^{-\beta(J-\frac{5}{2}J_1)}\!-\! 5{\rm e}^{-\beta(-\frac{J}{2}+J_1)}\right.
\nonumber\\
\!&\hspace*{5.5cm}-\!\left.35{\rm e}^{-\beta(-\frac{J}{2}-J_1)}\right]
\nonumber\\
\!&+\!\sinh\left(\frac{\beta h}{2}\right){\rm e}^{\frac{\beta}{2}h }\left[25{\rm e}^{-\beta(J-\frac{J_1}{2})}\!+\!3{\rm e}^{-\beta(J+\frac{5}{2}J_1)}\!+\!17{\rm e}^{-\beta(J-\frac{5}{2}J_1)}\!-\! 25{\rm e}^{-\beta(-\frac{J}{2}+J_1)}\!-\!55{\rm e}^{-\beta(-\frac{J}{2}-J_1)}\right]
\nonumber\\
\!&-\!10{\rm e}^{-\frac{\beta}{2}(-J-3J_1)}
\sinh\left(\frac{\beta}{2}(3J\!-\!4J_1)\right)\!-\!5{\rm e}^{2\beta J}\left[7\cosh\left(\frac{\beta}{2} J_1\right)\!+\!\sinh\left(\frac{\beta}{2} J_1\right)\right]\bigg\},
\label{d31}\\
\rho_{6,12}\!&=\rho_{12,6}\!=\!\!\rho_{7,13}(-h),
\label{d32}\\
\rho_{3,5}\!&=\!\rho_{5,3}\!=\!\frac{{\rm e}^{ \frac{\beta }{4}J_1}}{30{\cal Z}}\bigg\{
\cosh\left(\frac{\beta h}{2}\right){\rm e}^{\frac{\beta }{2} h}\left[7{\rm e}^{-\beta(J+\frac{5}{2}J_1)}\!-\!7{\rm e}^{-\beta(J-\frac{5}{2}J_1)}\right]
\nonumber\\
\!&+\!\sinh\left(\frac{\beta h}{2}\right){\rm e}^{\frac{\beta }{2} h}\left[{\rm e}^{-\beta(J+\frac{5}{2}J_1)}\!-\!{\rm e}^{-\beta(J-\frac{5}{2}J_1)}\right]
\nonumber\\
\!&-\!10{\rm e}^{-\beta(J-2J_1)}\sinh\left(\frac{3}{2}\beta J_1\right)
\!-\!10{\rm e}^{\frac{\beta}{2} J}{\rm e}^{\beta h}\sinh\left(\beta J_1\right)\bigg\},
\label{d33}\\
\rho_{14,16}\!&=\rho_{16,14}\!=\!\!\rho_{3,5}(-h),
\label{d34}\\
\rho_{3,11}\!&=\!\rho_{11,3}\!=\!\frac{\sqrt{2}}{2}\rho_{3,5},
\label{d35}\\
\rho_{8,16}\!&=\!\rho_{16,8}\!=\!\frac{\sqrt{2}}{2}\rho_{3,5}(-h),
\label{d36}\\
\rho_{5,11}\!&=\!\rho_{11,5}\!=\!\frac{\sqrt{2}{\rm e}^{ \frac{\beta }{4}J_1}}{180{\cal Z}}\bigg\{
\cosh\left(\frac{\beta h}{2}\right){\rm e}^{\frac{\beta }{2} h}\left[-30{\rm e}^{-\beta(-2J+\frac
{J_1}{2})}\!+\!42{\rm e}^{-\beta(J+\frac{5}{2}J_1)}\!+\!18{\rm e}^{-\beta(J-\frac{5}{2}J_1)}\right.
\nonumber\\
\!&\hspace*{6cm}-\! 25{\rm e}^{-\beta(-\frac{J}{2}+J_1)}\!-\!\left.15{\rm e}^{-\beta(-\frac{J}{2}-J_1)}\right]
\nonumber\\
\!&+\!\sinh\left(\frac{\beta h}{2}\right){\rm e}^{\frac{\beta }{2} h}\left[-10{\rm e}^{-\beta(-2J+\frac{J_1}{2})}\!+\!6{\rm e}^{-\beta(J+\frac{5}{2}J_1)}\!+\!14{\rm e}^{-\beta(J-\frac{5}{2}J_1)}\!-\! 5{\rm e}^{-\beta(-\frac{J}{2}+J_1)}\!+\!5{\rm e}^{-\beta(-\frac{J}{2}-J_1)}\right]
\nonumber\\
\!&+\!5{\rm e}^{-\beta(-\frac{J}{2}-2J_1)}
\left[\cosh\left(\frac{3\beta}{2}(J\!-\!J_1)\right)\!+\!3\sinh\left(\frac{3\beta}{2}(J\!-\!J_1)\right)\right]
\nonumber\\
\!&+\!5{\rm e}^{-\beta(J-2J_1)}
\left[3\cosh\left(\frac{3}{2}\beta J_1\right)\!-\!\sinh\left(\frac{3}{2}\beta J_1\right)\right]\bigg\},
\label{d37}\\
\rho_{8,14}\!&=\rho_{14,8}\!=\!\!\rho_{5,11}(-h),
\label{d38}\\
\rho_{5,7}\!&=\!\rho_{7,5}\!=\!\rho_{11,13}\!=\!\rho_{13,11}\!=\!\frac{{\rm e}^{ \frac{\beta }{4}J_1}}{90{\cal Z}}\bigg\{
\cosh\left(\frac{\beta h}{2}\right){\rm e}^{\frac{\beta }{2} h}\left[21{\rm e}^{-\beta(J+\frac{5}{2}J_1)}\!-\!11{\rm e}^{-\beta(J-\frac{5}{2}J_1)}\!+\! 35{\rm e}^{-\beta(-\frac{J}{2}+J_1)}\right.
\nonumber\\
&\!\hspace*{5cm}-\!\left.35{\rm e}^{-\beta(-\frac{J}{2}-J_1)}\right]
\nonumber\\
\!&+\!\sinh\left(\frac{\beta h}{2}\right){\rm e}^{\frac{\beta }{2} h}\left[3{\rm e}^{-\beta(J+\frac{5}{2}J_1)}\!-\!13{\rm e}^{-\beta(J-\frac{5}{2}J_1)}\!-\! 5{\rm e}^{-\beta(-\frac{J}{2}+J_1)}\!+\!5{\rm e}^{-\beta(-\frac{J}{2}-J_1)}\right]
\nonumber\\
\!&-\!20{\rm e}^{2\beta J}
\sinh\left(\frac{\beta}{2} J_1\right)\!-\!10{\rm e}^{-\beta(J-2J_1)}
\cosh\left(\frac{3}{2}\beta J_1\right)\bigg\},
\label{d39}\\
\rho_{12,14}\!&=\rho_{14,12}\!=\!\rho_{6,8}\!=\!\rho_{8,6}\!=\!\!\rho_{5,7}(-h),
\label{d40}\\
\rho_{5,13}\!&=\!\rho_{13,5}\!=\!\frac{\sqrt{2}{\rm e}^{ \frac{\beta }{4}J_1}}{180{\cal Z}}\bigg\{
\cosh\left(\frac{\beta h}{2}\right){\rm e}^{\frac{\beta }{2} h}\left[42{\rm e}^{-\beta(J+\frac{5}{2}J_1)}\!-\!22{\rm e}^{-\beta(J-\frac{5}{2}J_1)}\!-\! 5{\rm e}^{-\beta(-\frac{J}{2}+J_1)}\!+\!5{\rm e}^{-\beta(-\frac{J}{2}-J_1)}\right]
\nonumber\\
\!&+\!\sinh\left(\frac{\beta h}{2}\right){\rm e}^{\frac{\beta }{2} h}\left[6{\rm e}^{-\beta(J+\frac{5}{2}J_1)}\!-\!26{\rm e}^{-\beta(J-\frac{5}{2}J_1)}\!-\! 25{\rm e}^{-\beta(-\frac{J}{2}+J_1)}\!+\!25{\rm e}^{-\beta(-\frac{J}{2}-J_1)}\right]
\nonumber\\
\!&+\!20{\rm e}^{2\beta J}
\sinh\left(\frac{\beta}{2} J_1\right)\!-\!20{\rm e}^{-\beta(J-2J_1)}
\cosh\left(\frac{3}{2}\beta J_1\right)\bigg\},
\label{d41}\\
\rho_{6,14}\!&=\rho_{14,6}\!=\!\!\rho_{5,13}(-h),
\label{d42}\\
\rho_{7,11}\!&=\!\rho_{11,7}\!=\!\frac{\sqrt{2}{\rm e}^{ \frac{\beta }{4}J_1}}{180{\cal Z}}\bigg\{
\cosh\left(\frac{\beta h}{2}\right){\rm e}^{\frac{\beta }{2} h}\left[21{\rm e}^{-\beta(J+\frac{5}{2}J_1)}\!-\!11{\rm e}^{-\beta(J-\frac{5}{2}J_1)}\!-\! 40{\rm e}^{-\beta(-\frac{J}{2}+J_1)}\!+\!40{\rm e}^{-\beta(-\frac{J}{2}-J_1)}\right]
\nonumber\\
\!&+\!\sinh\left(\frac{\beta h}{2}\right){\rm e}^{\frac{\beta }{2} h}\left[3{\rm e}^{-\beta(J+\frac{5}{2}J_1)}\!-\!13{\rm e}^{-\beta(J-\frac{5}{2}J_1)}\!-\! 20{\rm e}^{-\beta(-\frac{J}{2}+J_1)}\!+\!20{\rm e}^{-\beta(-\frac{J}{2}-J_1)}\right]
\nonumber\\
\!&+\!40{\rm e}^{2\beta J}
\sinh\left(\frac{\beta}{2} J_1\right)\!-\!10{\rm e}^{-\beta(J-2J_1)}
\cosh\left(\frac{3}{2}\beta J_1\right)\bigg\},
\label{d43}\\
\rho_{8,12}\!&=\rho_{12,8}\!=\!\!\rho_{7,11}(-h).
\label{d44}
\end{flalign}
\end{widetext}
Subsequently, the partial transposition of the reduced density matrix  $\hat{\rho}_{\mu_{1}|S_{1}S_{2}}^{T_{\mu_{1}}}$,   with respect the  spin $\mu_{1}$, has the following block-diagonal structure involving two $1\times 1$ matrices $\mathbf{R}_1^{\mu_{1}}$, $\overline{\mathbf{R}}_1^{\mu_{1}}\!=\!\mathbf{R}_1^{\mu_{1}}(-h)$,  two $3\times 3$ matrices $\mathbf{R}_2^{\mu_1}$, $\overline{\mathbf{R}}_2^{\mu_{1}}\!=\!\mathbf{R}_2^{\mu_{1}}(-h)$ and two $5\times 5$ matrices $\mathbf{R}_3^{\mu_1}$, $\overline{\mathbf{R}}_3^{\mu_{1}}\!=\!\mathbf{R}_3^{\mu_{1}}(-h)$ with the following elements
\begin{align}
\allowdisplaybreaks
\mathbf{R}_1^{\mu_{1}}\!&=\!\left(
\begin{array}{c}
 \rho_{9,9} \\
\end{array}\right),\mathbf{R}_2^{\mu_{1}}\!=\!\left(
\begin{array}{ccc}
 \rho_{1,1} & \rho_{2,10} & \rho_{4,10}\\
 \rho_{2,10} & \rho_{11,11} & \rho_{11,13}\\
 \rho_{4,10} & \rho_{11,13} & \rho_{13,13}\\
\end{array}\right),\;\;
\nonumber\\
\mathbf{R}_3^{\mu_{1}}\!&=\!\left(
\begin{array}{ccccc}
 \rho_{2,2} & \rho_{2,4} & \rho_{3,11}& \rho_{5,11}& \rho_{7,11}\\
 \rho_{2,4} & \rho_{4,4} & \rho_{3,13}& \rho_{5,13}& \rho_{7,13}\\
 \rho_{3,11} & \rho_{3,13} & \rho_{12,12}& \rho_{12,14}& \rho_{12,16}\\
  \rho_{5,11} & \rho_{5,13} & \rho_{12,14}& \rho_{14,14}& \rho_{14,16}\\
 \rho_{7,11} & \rho_{7,13} & \rho_{12,16}& \rho_{14,16}& \rho_{16,16} 
\end{array}\right).
\label{d45}
\end{align}
Due to the complexity of all diagonal blocks, the respective eigenvalues were calculated numerically.  The sum of absolute values of all negative eigenvalues of $\mathbf{R}_2^{\mu_{1}}$, $\overline{\mathbf{R}}_2^{\mu_{1}}$, $\mathbf{R}_3^{\mu_{1}}$ and  $\overline{\mathbf{R}}_3^{\mu_{1}}$ determines the respective bipartite negativity ${\cal N}_{\mu_1|S_1S_2}$ between a single spin $\mu_1$ and the spin dimer $S_1-S_2$.

\section{Bipartite entanglement ${\cal N}_{S_{1}|\mu_{1}S_{2}}$ between the single spin $S_1$ and the spin dimer $\mu_1-S_2$}
\label{App E}
The reduced density operator $\hat{\rho}_{S_{1}|\mu_{1}S_{2}}$ is identical with the reduced density operator $\hat{\rho}_{\mu_{1}|S_{1}S_{2}}$ (Eq.~\eqref{d1}) and of course, the corresponding reduced density matrix is the same as that presented in Eq.~\eqref{d2}. A partial transpose of the density matrix~\eqref{d2} over a different spin, namely the spin $S_{1}$, leads again to the  block-diagonal structure with   two $1\times 1$ matrices $\mathbf{R}_1^{S_{1}}$, $\overline{\mathbf{R}}_1^{S_{1}}\!=\!\mathbf{R}_1^{S_{1}}(-h)$, two $3\times 3$ matrices $\mathbf{R}_2^{S_{1}}$, $\overline{\mathbf{R}}_2^{S_{1}}\!=\!\mathbf{R}_2^{S_{1}}(-h)$ and  two $5\times 5$ matrices $\mathbf{R}_3^{S_{1}}$, $\overline{\mathbf{R}}_3^{S_{1}}\!=\!\mathbf{R}_3^{S_{1}}(-h)$. Nevertheless,  the structure of the each block matrix is different. In particular,
\begin{align}
\allowdisplaybreaks
\mathbf{R}_1^{S_{1}}\!&=\!\left(
\begin{array}{c}
 \rho_{7,7} \\
\end{array}\right),
\mathbf{R}_2^{S_{1}}\!=\!\left(
\begin{array}{ccc}
 \rho_{3,3} & \rho_{3,11} & \rho_{6,12}\\
 \rho_{3,11} & \rho_{11,11} & \rho_{12,14}\\
 \rho_{6,12} & \rho_{12,14} & \rho_{15,15}\\
\end{array}\right),\;\;
\nonumber\\
\mathbf{R}_3^{S_{1}}\!&=\!\left(
\begin{array}{ccccc}
 \rho_{1,1} & \rho_{2,4} & \rho_{4,10}& \rho_{3,7}& \rho_{7,11}\\
 \rho_{2,4} & \rho_{5,5} & \rho_{5,13}& \rho_{6,8}& \rho_{8,14}\\
 \rho_{4,10} & \rho_{5,13} & \rho_{13,13}& \rho_{6,16}& \rho_{14,16}\\
  \rho_{3,7} & \rho_{6,8} & \rho_{6,16}& \rho_{9,9}& \rho_{9,17}\\
 \rho_{7,11} & \rho_{8,14} & \rho_{14,16}& \rho_{9,17}& \rho_{17,17} 
\end{array}\right).
\label{e1}
\end{align}
Particular elements $\rho_{i,j}$ are the same as that presented in Eqs.~\eqref{d3}-\eqref{d44}. The sum of absolute values of all negative eigenvalues of $\mathbf{R}_2^{S_{1}}$, $\overline{\mathbf{R}}_2^{S_{1}}$,  $\mathbf{R}_3^{S_{1}}$, $\overline{\mathbf{R}}_3^{S_{1}}$ determines the respective bipartite negativity ${\cal N}_{S_1|\mu_1S_2}$ between a single spin $S_1$ and the spin dimer $\mu_1-S_2$.

\section{Bipartite entanglement ${\cal N}_{S_{2}|\mu_{1}S_{1}}$ between the single spin $S_2$ and the spin dimer $\mu_1-S_1$}
\label{App F}
The reduced density operator $\hat{\rho}_{S_{2}|\mu_{1}S_{1}}$ is identical with the reduced density operator $\hat{\rho}_{\mu_{1}|S_{1}S_{2}}$ (Eq.~\eqref{d1}) and of course, the corresponding reduced density matrix is the same as that presented in Eq.~\eqref{d2}. A partial transpose of the density  matrix~\eqref{d2} over a different spin, namely the spin $S_{2}$, leads again to the  block-diagonal structure with    two $1\times 1$ matrices $\mathbf{R}_1^{S_{2}}$, $\overline{\mathbf{R}}_1^{S_{2}}\!=\!\mathbf{R}_1^{S_{2}}(-h)$, two $3\times 3$ matrices $\mathbf{R}_2^{S_{2}}$, $\overline{\mathbf{R}}_2^{S_{2}}\!=\!\mathbf{R}_2^{S_{2}}(-h)$ and  two $5\times 5$ matrices $\mathbf{R}_3^{S_{2}}$, $\overline{\mathbf{R}}_3^{S_{2}}\!=\!\mathbf{R}_3^{S_{2}}(-h)$. Nevertheless,  the structure of the each block matrix is different. In particular,
\begin{align}
\allowdisplaybreaks
\mathbf{R}_1^{S_{2}}\!&=\!\left(
\begin{array}{c}
 \rho_{3,3} \\
\end{array}\right),
\mathbf{R}_2^{S_{2}}\!=\!\left(
\begin{array}{ccc}
 \rho_{2,2} & \rho_{3,5} & \rho_{3,11}\\
 \rho_{3,5} & \rho_{6,6} & \rho_{6,12}\\
 \rho_{3,11} & \rho_{6,12} & \rho_{12,12}\\
\end{array}\right),\;\;
\nonumber\\
\mathbf{R}_3^{S_{2}}\!&=\!\left(
\begin{array}{ccccc}
 \rho_{1,1} & \rho_{2,4} & \rho_{2,10}& \rho_{3,7}& \rho_{3,13}\\
 \rho_{2,4} & \rho_{5,5} & \rho_{5,11}& \rho_{6,8}& \rho_{6,14}\\
 \rho_{2,10} & \rho_{5,11} & \rho_{11,11}& \rho_{8,12}& \rho_{12,14}\\
  \rho_{3,7} & \rho_{6,8} & \rho_{8,12}& \rho_{9,9}& \rho_{9,15}\\
 \rho_{3,13} & \rho_{6,14} & \rho_{12,14}& \rho_{9,15}& \rho_{15,15} 
\end{array}\right).
\label{k1}
\end{align}
Particular elements $\rho_{i,j}$ are the same as that presented in Eqs.~\eqref{d3}-\eqref{d44}. The sum of absolute values of all negative eigenvalues of $\mathbf{R}_2^{S_{2}}$, $\overline{\mathbf{R}}_2^{S_{2}}$,  $\mathbf{R}_3^{S_{2}}$, $\overline{\mathbf{R}}_3^{S_{2}}$ determines the respective bipartite negativity ${\cal N}_{S_2|\mu_1S_1}$ between a single spin $S_2$ and the spin dimer $\mu_1-S_1$.\\

\end{document}